\documentclass[zpreprint,zbstepj]{zeus_paper}
\usepackage{lineno}
\usepackage[english]{babel}

\newcommand{\ZcoosysB}{%
The ZEUS coordinate system is a right-handed Cartesian system, with the $Z$
axis pointing in the proton beam direction, referred to as the ``forward
direction'', and the $X$ axis pointing left towards the centre of HERA.
The coordinate origin is at the nominal interaction point.\xspace}

\newcommand{\ZcoosysfnB}{\footnote{\ZcoosysB}}

\chardef\usc=95
\chardef\til=126
\catcode`\@=11 
\DeclareRobustCommand\xdotspace{\futurelet\@let@token\@xdotspace}
\def\@xdotspace{%
  \ifx\@let@token.\else
  \ifx\@let@token\bgroup.\else
  \ifx\@let@token\egroup.\else
  \ifx\@let@token\/.\else
  \ifx\@let@token\ .\else
  \ifx\@let@token~.\else
  \ifx\@let@token!.\else
  \ifx\@let@token,.\else
  \ifx\@let@token:.\else
  \ifx\@let@token;.\else
  \ifx\@let@token?.\else
  \ifx\@let@token/.\else
  \ifx\@let@token'.\else
  \ifx\@let@token).\else
  \ifx\@let@token-.\else
  \ifx\@let@token\@xobeysp.\else
  \ifx\@let@token\space.\else
  \ifx\@let@token\@sptoken.\else
   .\space
   \fi\fi\fi\fi\fi\fi\fi\fi\fi\fi\fi\fi\fi\fi\fi\fi\fi\fi}
\catcode`\@=12 

\newcommand{\stru}[2]{%
   \relax\ifmmode\hbox{\vrule height#1 depth#2 width0pt}%
   \else\vrule height#1 depth#2 width0pt\fi}

\newcommand{\Ronum}[1]{\uppercase\expandafter{\romannumeral#1}}
\newcommand{\ronum}[1]{\expandafter{\romannumeral#1}}
\DeclareRobustCommand{\LaTeXZ}{%
  \LaTeX\kern-.05em4\kern-.1em
  {\raisebox{-0.2ex}{$\scriptstyle\text{ZEUS}$}}\xspace}

\newcommand{\ouml}{\"o}
\newcommand{\uuml}{\"u}

\DeclareMathAlphabet{\mathbf}{OT1}{cmr}{bx}{sl}
\newcommand{\eVdist}{\kern-0.06667em}

\newcommand{\Gev}{{\text{Ge}\eVdist\text{V\/}}}

\newcommand{\mev}{{\,\text{Me}\eVdist\text{V\/}}}
\newcommand{\gev}{{\,\text{Ge}\eVdist\text{V\/}}}

\newcommand{\pbi}{\,\text{pb}^{-1}}

\newcommand{\met}{\,\text{m}}

\newcommand{\mum}{\,\upmu\text{m}}
\newcommand{\mm}{\,\text{mm}}
\newcommand{\cm}{\,\text{cm}}

\newcommand{\Tesla}{\,\text{T}}

\newcommand{\slashfrac}[2]{%
  \raisebox{0.5ex}{\ensuremath #1}\kern-0.12em/\kern-0.08em
  \raisebox{-.8ex}{\ensuremath #2}}

\newcommand{\sqr}[3]{%
    {\vcenter{\hrule height.#3ex\hbox{\vrule width.#2ex height#1ex
     \kern#1ex\vrule width.#3ex}\hrule height.#2ex}}}

\newcommand{\widebar}[1]{%
   \mkern1.5mu\overline{\mkern-1.5mu#1\mkern-1.mu}\mkern1.mu}
\catcode`\@=11 
\newcommand{\parenbar}{\mathpalette\p@renb@r}
\def\p@renb@r#1#2{\vbox{%
  \ifx#1\scriptscriptstyle \dimen@.7em\dimen@ii.2em\else
  \ifx#1\scriptstyle \dimen@.8em\dimen@ii.25em\else
  \dimen@1em\dimen@ii.4em\fi\fi \offinterlineskip
  \ialign{\hfill##\hfill\cr
    \vbox{\hrule width\dimen@ii}\cr
    \noalign{\vskip-.3ex}%
    \hbox to\dimen@{$\mathchar300\hfil\mathchar301$}\cr
    \noalign{\vskip-.3ex}%
    $#1#2$\cr}}}
\catcode`\@=12 

\newcommand{\qbar}{\widebar{q}}

\newcommand{\sihat}{{\hat\sigma}}

\newcommand{\IP}{{\rm I$\kern-0.01667em$P}\xspace}

\newcommand{\Lumi}{{\cal L}}

\mathchardef\qsm=63
\mathchardef\pls=43
\mathchardef\mns=512
\mathchardef\plm=518
\mathchardef\eql=61
\mathchardef\smallleft=300
\mathchardef\smallright=301
\mathchardef\les=316
\mathchardef\gre=318
\mathchardef\leq=532
\mathchardef\grq=533

\catcode`\@=11 
\newcounter{pict@width}
\newcounter{pict@height}
\newlength{\pict@scale}
\newcommand{\psfigadd}[4]{%
\setcounter{pict@width}{1*\ratio{#2+\pict@scale/2}{\pict@scale}}
\setcounter{pict@height}{1*\ratio{#3+\pict@scale/2}{\pict@scale}}
\setlength{\unitlength}{\pict@scale}
\hbox to #2{\hspace{-\fill}\begin{picture}(\thepict@width,\thepict@height)
\put(0,0){\psfig{figure=#1,width=#2,height=#3,clip=}}
\SetScale{0.283466457}
\SetWidth{1.763889}
{#4}
\end{picture}}
}
\newcounter{pict@widthfst}
\newcounter{pict@widthscd}
\newcounter{pict@widthtot}
\newcommand{\psfigaddtwo}[7]{%
\setcounter{pict@widthfst}{1*\ratio{#2+\pict@scale/2}{\pict@scale}}
\setcounter{pict@widthscd}{1*\ratio{#2+#4+\pict@scale/2}{\pict@scale}}
\setcounter{pict@widthtot}{1*\ratio{#2+#4+#6+\pict@scale/2}{\pict@scale}}
\setcounter{pict@height}{1*\ratio{#3+\pict@scale/2}{\pict@scale}}
\setlength{\unitlength}{\pict@scale}
\hbox{\hspace{-\fill}\begin{picture}(\thepict@widthtot,\thepict@height)
\put(0,0){\psfig{figure=#1,width=#2,height=#3,clip=}}
\put(\thepict@widthscd,0){\psfig{figure=#5,width=#6,height=#3,clip=}}
\SetScale{0.283466457}
\SetWidth{1.763889}
{#7}
\end{picture}}
}
\newcommand{\psfigror}[4]{%
\setcounter{pict@width}{1*\ratio{#2+\pict@scale/2}{\pict@scale}}
\setcounter{pict@height}{1*\ratio{#3+\pict@scale/2}{\pict@scale}}
\setlength{\unitlength}{\pict@scale}
\hbox{\begin{picture}(\thepict@width,\thepict@height)
\put(0,\thepict@height){\psfig{figure=#1,width=#3,height=#2,clip=,angle=270}}
\SetScale{0.283466457}
\SetWidth{1.763889}
{#4}
\end{picture}}
}
\newcommand{\psfigrol}[4]{%
\setcounter{pict@width}{1*\ratio{#2+\pict@scale/2}{\pict@scale}}
\setcounter{pict@height}{1*\ratio{#3+\pict@scale/2}{\pict@scale}}
\setlength{\unitlength}{\pict@scale}
\hbox{\begin{picture}(\thepict@width,\thepict@height)
\put(0,0){\psfig{figure=#1,width=#3,height=#2,clip=,angle=90}}
\SetScale{0.283466457}
\SetWidth{1.763889}
{#4}
\end{picture}}
}
\catcode`\@=12 
\newlength\listtextwidth

\catcode`\@=11 
\newlength{\@tabfninsert}
\newlength{\@tabfnwidth}
\newcommand{\tabfootnote}[2]{%
  \setlength{\@tabfninsert}{0.8em}
  \setlength{\@tabfnwidth}{\textwidth}
  \addtolength{\@tabfnwidth}{-\@tabfninsert}
  \addtolength{\@tabfnwidth}{-0.4em}
  \noindent\makebox[\@tabfninsert][r]{\footnotesize$^{#1}$\hfil}\hfill%
  \parbox[t]{\@tabfnwidth}{\footnotesize #2\hfill}}
\catcode`\@=12 

\begin{document}


\prepnum{{DESY--07--126}}

\title{
  Dijet production in\\
  diffractive deep inelastic scattering\\
  at HERA
}                                                       
                   
\author{ZEUS Collaboration}
\date{August 2007}

\abstract{ The production of dijets in diffractive deep inelastic
scattering has been measured with the ZEUS detector at HERA using an
integrated luminosity of $61 \pbi$. The dijet cross section has been
measured for virtualities of the exchanged virtual photon, $5 < Q^2 <
100 \gev^2$, and $\gamma^{*} p$ centre-of-mass energies, $100 < W <
250$ GeV. The jets, identified using the inclusive $k_{T}$ algorithm
in the $\gamma^* p$ frame, were required to have a transverse energy
$E^*_{T, \rm jet} > 4 \gev$ and the jet with the highest transverse
energy was required to have $E^*_{T,\rm jet} > 5 \gev$.  All jets were
required to be in the pseudorapidity range $-3.5 < \eta^*_{\rm jet } <
0$.  The differential cross sections are compared to leading-order
predictions and next-to-leading-order QCD calculations based on recent
diffractive parton densities extracted from inclusive diffractive deep
inelastic scattering data.  }

\makezeustitle

\begin{center}                                                                                     
{                      \Large  The ZEUS Collaboration              }                               
\end{center}                                                                                       
  S.~Chekanov$^{   1}$,                                                                            
  M.~Derrick,                                                                                      
  S.~Magill,                                                                                       
  B.~Musgrave,                                                                                     
  D.~Nicholass$^{   2}$,                                                                           
  \mbox{J.~Repond},                                                                                
  R.~Yoshida\\                                                                                     
 {\it Argonne National Laboratory, Argonne, Illinois 60439-4815, USA}~$^{n}$                       
\par \filbreak                                                                                     
  M.C.K.~Mattingly \\                                                                              
 {\it Andrews University, Berrien Springs, Michigan 49104-0380, USA}                               
\par \filbreak                                                                                     
  M.~Jechow, N.~Pavel~$^{\dagger}$, A.G.~Yag\"ues Molina \\                                        
  {\it Institut f\"ur Physik der Humboldt-Universit\"at zu Berlin,                                 
           Berlin, Germany}~$^{b}$
% ...
\newpage                                                                                           
$^{\    1}$ supported by DESY, Germany \\                                                          
$^{\    2}$ also affiliated with University College London, UK \\                                  
% ...
$^{\dagger}$ deceased \\                                                                           
%
                                                           %
                                                           %
\begin{tabular}[h]{rp{14cm}}                                                                       
$^{a}$ &  supported by the Natural Sciences and Engineering Research Council of Canada (NSERC) \\  
$^{b}$ &  supported by the German Federal Ministry for Education and Research (BMBF), under        
          contract numbers 05 HZ6PDA, 05 HZ6GUA, 05 HZ6VFA and 05 HZ4KHA\\                         
$^{c}$ &  supported in part by the MINERVA Gesellschaft f\"ur Forschung GmbH, the Israel Science   
          Foundation (grant no. 293/02-11.2) and the U.S.-Israel Binational Science Foundation \\  
$^{d}$ &  supported by the German-Israeli Foundation and the Israel Science Foundation\\           
$^{e}$ &  supported by the Italian National Institute for Nuclear Physics (INFN) \\                
$^{f}$ &  supported by the Japanese Ministry of Education, Culture, Sports, Science and Technology 
          (MEXT) and its grants for Scientific Research\\                                          
$^{g}$ &  supported by the Korean Ministry of Education and Korea Science and Engineering          
          Foundation\\                                                                             
$^{h}$ &  supported by the Netherlands Foundation for Research on Matter (FOM)\\                   
$^{i}$ &  supported by the Polish State Committee for Scientific Research, grant no.               
          620/E-77/SPB/DESY/P-03/DZ 117/2003-2005 and grant no. 1P03B07427/2004-2006\\             
$^{j}$ &  partially supported by the German Federal Ministry for Education and Research (BMBF)\\   
$^{k}$ &  supported by RF Presidential grant N 8122.2006.2 for the leading                         
          scientific schools and by the Russian Ministry of Education and Science through its grant
          Research on High Energy Physics\\                                                        
$^{l}$ &  supported by the Spanish Ministry of Education and Science through funds provided by     
          CICYT\\                                                                                  
$^{m}$ &  supported by the Particle Physics and Astronomy Research Council, UK\\                   
$^{n}$ &  supported by the US Department of Energy\\                                               
$^{o}$ &  supported by the US National Science Foundation. Any opinion,                            
findings and conclusions or recommendations expressed in this material                             
are those of the authors and do not necessarily reflect the views of the                           
National Science Foundation.\\                                                                     
$^{p}$ &  supported by the Polish Ministry of Science and Higher Education                         
as a scientific project (2006-2008)\\                                                              
$^{q}$ &  supported by FNRS and its associated funds (IISN and FRIA) and by an Inter-University    
          Attraction Poles Programme subsidised by the Belgian Federal Science Policy Office\\     
$^{r}$ &  supported by the Malaysian Ministry of Science, Technology and                           
Innovation/Akademi Sains Malaysia grant SAGA 66-02-03-0048\\                                       
\end{tabular}                                                                                      
                                                           %
                                                           %

\newcommand {\pom} {I\!\!P}
\newcommand {\pomsub} {{\scriptscriptstyle \pom}}
\newcommand{\xpom}{\ensuremath{x_{\pomsub}}}
\newcommand{\xpomobs}{\ensuremath{x^{\rm obs}_{\pomsub}}}
\newcommand{\zpomobs}{\ensuremath{z^{\rm obs}_{\pomsub}}}
\newcommand{\xgammaobs}{\ensuremath{x^{\rm obs}_{\gamma}}}
\newcommand{\zpom}{\ensuremath{z_{\pomsub}}}
\newcommand{\xgamma}{\ensuremath{x_{\gamma}}}
\newcommand{\etjets}{\ensuremath{E_{\rm T,jet}^{*}}}
\newcommand{\etajets}{\ensuremath{\eta^{*}_{\rm jet}}}
\newcommand{\etjj}{\ensuremath{E_{\rm T,J}^{*}}}
\newcommand{\etajj}{\ensuremath{\eta^{*}_{\rm J}}}
\newcommand{\honefitOLD}{``H1 2002 fit (prel.)''}
\newcommand{\honefitA}{{\rm H1~2006 - Fit A}}
\newcommand{\honefitB}{{\rm H1~2006 - Fit B}}

\pagenumbering{arabic} 
\pagestyle{plain}
\section{Introduction}
\label{sec-int}

Diffractive events in deep inelastic scattering (DIS) are characterised by the
presence of a fast forward proton, a large rapidity gap (LRG) - an
angular region between the scattered proton and the dissociated photon
with no particle flow \cite{ZEUSDiff94,ZEUSDiffBPC,ZEUSMx,H1fit2,
  H1fit2006, LPSfit} - and a dissociated virtual photon $\gamma^*$. In
recent years perturbative QCD (pQCD) has become a successful tool for
describing diffractive events \cite{H1fit2, H1fit2006, LPSfit,MRWfit}.
The cross section for diffractive DIS
processes can be described by a convolution of universal diffractive
parton distribution functions (dPDFs) and process-dependent
coefficients, which can be calculated in pQCD \cite{Collins}.  At
HERA, dPDFs have been determined using inclusive diffractive DIS data
\cite{H1fit2, H1fit2006, LPSfit}.

\begin{figure}[phtb]
\begin{center}\includegraphics[%
  height=7cm]{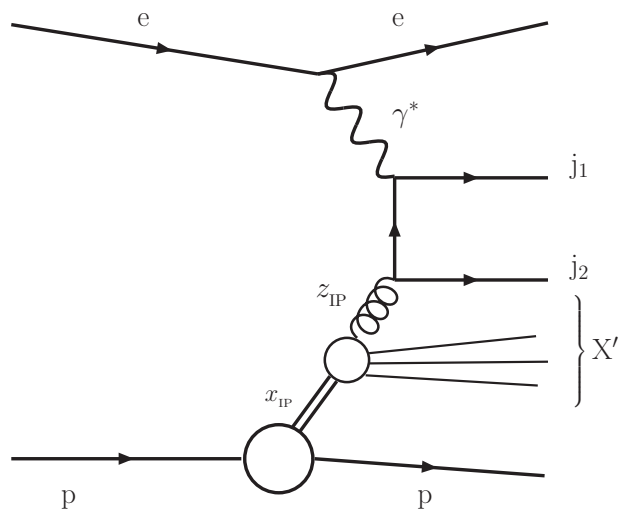}
\end{center}

\caption{\label{cap:BGF_diagram} Schematic representation of the boson-gluon fusion diagram for LO dijet production in diffractive DIS.
}
\end{figure}

This paper presents measurements of dijet production in diffractive
neutral current DIS with the ZEUS detector at HERA. The presence of a
hard scale in such a process, either the virtuality of the photon or
the large jet transverse momentum, is well suited for a pQCD analysis.
Dijet processes are particularly sensitive to the density of gluons in
the diffractive exchange (i.e.  via $\gamma^{*} g\rightarrow
q\bar{q}$, as shown in Fig.~\ref{cap:BGF_diagram}), and gluons have
been shown to carry most of the momentum of the colourless exchange
\cite{H1fit2,H1fit2006,ZEUSincldiffDIS}. The measured differential
cross sections are compared with leading-order (LO) and
next-to-leading-order (NLO) QCD predictions using the available dPDFs.
The results presented here benefit from higher statistics compared to
previous measurements of the same process \cite{H1Dijets}.

\section{Experimental set-up}
\label{sec-exp}

This analysis is based on  $61 \pbi$ of data collected with the ZEUS detector
at the HERA collider during the 1999-2000 data-taking
period.  During this period, HERA collided either electrons or
positrons\footnote{In the following, for simplicity, the word positron
will be used to denote both electrons and positrons. The integrated
luminosity for $e^{-}p$ data is $3\,\pbi$, while for $e^{+}p$ data is $58\,\pbi$.} of $27.5 \gev$ with protons
of $920 \gev$ at  a centre-of-mass energy of $\sqrt{s}=318
\gev$.

A detailed description of the ZEUS detector can be found 
elsewhere~\cite{ZEUSbluebook}. A brief outline of the 
components that are most relevant for this analysis is given
below.\xspace

Charged particles are tracked in the central tracking detector
(CTD)~\cite{CTDref1,*CTDref2,*CTDref3}, which operates
in a magnetic field of $1.43\Tesla$ provided by a thin superconducting
coil. The CTD consists of 72~cylindrical drift chamber layers,
organised in 9~superlayers covering the polar-angle region\ZcoosysfnB~
\mbox{$15^\circ<\theta<164^\circ$}. The transverse-momentum resolution for
full-length tracks is $\sigma(p_T)/p_T=0.0058p_T\oplus0.0065\oplus0.0014/p_T$,
with $p_T$ in $\Gev$.

The high-resolution uranium--scintillator calorimeter
(CAL)~\cite{nim:a309:77,*nim:a309:101,*nim:a321:356,*nim:a336:23} consists of
three parts: the forward (FCAL), the barrel (BCAL) and the rear (RCAL)
calorimeters. Each part is subdivided transversely into towers and
longitudinally into one electromagnetic section and either one
(in RCAL) or two (in BCAL and FCAL) hadronic sections. The
smallest subdivision of the calorimeter is called a cell.  The CAL
energy resolutions, as measured under test-beam conditions, are
$\sigma(E)/E=0.18/\sqrt{E}$ for electrons and
$\sigma(E)/E=0.35/\sqrt{E}$ for hadrons, with $E$ in $\Gev$.

During the 1999-2000 data-taking period, the forward plug calorimeter (FPC)
\cite{FPCdet}, located in the beam hole of FCAL, extended the
pseudorapidity coverage of the calorimeter up to values of $\eta
\approx 5$. It consisted of a lead-scintillator calorimeter read out
by wavelength shifters and photomultipliers.

In order to improve the detection of positrons scattered at low
angles, the angular coverage in the rear direction was extended by means of the
small rear tracking detector (SRTD) \cite{SRTDref1,SRTDref2}. The SRTD
consists of two planes of 1 cm wide and 0.5 cm thick scintillator
strips glued on the front of RCAL. The orientations of the strips in
the two planes are orthogonal. Scattered positrons were also detected
in the rear hadron-electron separator (RHES) \cite{HESref1}, a matrix
of more than 10000 silicon diodes $400 \mum$ thick inserted in the
RCAL.

The luminosity was measured using the bremsstrahlung process
$ep\rightarrow ep\gamma$ with the luminosity monitor
\cite{LUMIref1,*LUMIref2,*LUMIref3}, a lead-scintillator calorimeter
placed in the HERA tunnel at $Z=-107\met$.



\section{Kinematics}
\label{sec-kin}

Dijet production in diffractive DIS ($ep\rightarrow
e+p+{\rm j1+j2+X'}$) is characterised by the simultaneous presence
of a scattered positron, a scattered proton $p$ that escapes
undetected down the beam pipe, and the photon-dissociative system ${\rm X}$,
which contains the dijet system ${\rm j1 + j2}$, produced in the hard
scattering along with the rest of the hadronic system ${\rm X'}$ (see
Fig.~\ref{cap:BGF_diagram}).  Deep inelastic scattering of a positron
on a proton is described by the following kinematic variables:
\begin{itemize}
\item $s=(P+k)^{2}$, the squared $ep$ centre-of-mass energy, where $P$
  and $k$ indicate the incoming proton and the incoming positron
  four-momenta, respectively;
\item $Q^{2}=-q^{2}=-(k-k')^{2}$, the virtuality of $\gamma^{\ast}$,
  where $k'$ is the four-momentum of the scattered positron;
\item $W^{2}=(P+q)^{2}$, the  centre-of-mass energy squared of the $\gamma^{\ast}p$ system.
\end{itemize}

Diffractive events are further characterised by the variables:
\begin{itemize}
\item $M_{\rm X}$, the invariant mass of the photon-dissociative system;
\item $t=(P-P')^{2}$, the squared four-momentum transfer at the proton
  vertex, where $P'$ denotes the four-momentum of the scattered proton;
\item $\xpom= (P-P') \cdot q/ P \cdot q$, the momentum fraction lost by the proton; 
\item $\beta=Q^{2} / 2(P-P')\cdot q$, a measure of the fractional momentum of the diffractive
exchange carried by the struck parton.
\end{itemize}

The description of the dijet system in the hadronic final state requires the use of additional variables:
\begin{itemize}
\item $\zpom$, the fraction of the
momentum of the diffractive exchange carried by the parton
participating in the hard process and defined as
\begin{equation}\label{eq-zpom}
\zpom={\displaystyle \frac{q\cdot v}{q\cdot (P-P')}},
\end{equation}
 where $v$ is the four-momentum of the parton originating from the diffractive exchange;
\item $\xgamma$, the fractional momentum of the virtual
  photon participating in the hard process. In DIS, $\xgamma$ is expected
  to be unity (direct photon).  However, some models introduce the
  concept of a resolved virtual photon, where the $\gamma^{\ast}$
  can fluctuate into a partonic state before participating in the hard
  interaction. For resolved photon processes, $\xgamma$ is expected to be
  lower than unity.  The variable $\xgamma$ is defined as
\begin{equation}\label{eq-xgam}
\xgamma={\displaystyle \frac{P\cdot u}{P\cdot q}},
\end{equation}
where $u$ is the four-momentum of the parton originating from the
virtual photon. 
\end{itemize}

\section{Theoretical models}
\label{sec-mod}

\subsection{QCD factorisation in diffraction}
\label{sec-mod-qcdfact}

The cross section for diffractive DIS processes at fixed $s$ depends
in general on four independent variables, which are usually chosen to
be $Q^{2}, \beta,\xpom$ and $t$. According to the QCD factorisation theorem
\cite{Collins}, the cross section for inclusive diffraction,
$\sigma(\gamma^{*}p\rightarrow {\rm X}p)$, can be written as

\begin{equation}
\label{eq-fact1}
\frac{d^{2}\sigma}{d\xpom dt}={ \displaystyle \sum_{i=q,\qbar,g}\int dQ^{2}\int^{1}_{\beta}d\xi\,\sihat^{\gamma^{*}i}(Q^{2},\xi)f^{D}_{i}(\xpom,t,\xi,Q^{2})}.
\end{equation}

This expression is valid at fixed $\xpom$ and $t$ and for scales
sufficiently large to permit the use of pQCD. The sum runs over all
partons. The partonic cross-section
$\sihat^{\gamma^{*}i}(Q^{2},\beta)$ for the hard subprocess involving
the virtual photon and the parton $i$ is calculable in pQCD. The
functions $f^{D}_{i}(\xpom,t,\beta,Q^{2})$ are the dPDFs: they
describe the probability to find in the proton a parton of kind $i$
carrying a fraction $\xpom\cdot\beta$ of its momentum with a probe of
resolution $Q^{2}$ under the condition that the proton stays intact,
with a momentum loss quantified by $\xpom$ and $t$.
For diffractive production of dijets, Eq.~(\ref{eq-fact1}) is rewritten as:
\begin{displaymath}
\label{eq-fact2}
\frac{d^{2}\sigma_{jj}}{d\xpom dt}={ \displaystyle \sum_{i=q,\qbar,g}\int dQ^{2}\int^{1}_{z_{\mathbb P}}d\xi\,\sihat^{\gamma^{*}i}_{jj}(Q^{2},\xi)f^{D}_{i}(\xpom,t,\xi,Q^{2})},
\end{displaymath}

where now $\zpom$ is the variable sensitive to the dPDFs and the
subprocess cross section $\sigma^{\gamma^\ast i}$ is replaced by the cross
section, $\sigma^{\gamma^{*}i}_{jj}$, for the reaction $\gamma^{*}i\rightarrow {\rm j1\;j2}$.

At HERA, the dPDFs have been determined within the QCD DGLAP formalism
\cite{DGLAPref1,DGLAPref2,DGLAPref3,DGLAPref4} by means of fits to
inclusive diffractive DIS measurements with a procedure similar to
that used to extract the standard proton PDFs from inclusive DIS data
\cite{PDFfitref1,PDFfitref2,PDFfitref3,H1PDFref1,ZEUSPDFref1,ZEUSNLO-jets}.
Consistency between the measured cross sections for
semi-inclusive processes and calculations using these dPDFs represents
an experimental proof of the validity of the QCD factorization
hypothesis in diffraction \cite{H1Dijets,H1Dstar}.

Most of the dPDF parameterisations use Regge phenomenology arguments
\cite{ReggePhenom} to factorise the ($\xpom,t$) from the ($\beta,Q^{2}$)
dependence. In the Regge approach, diffractive scattering proceeds via
the exchange of the Pomeron trajectory. The dPDFs are then written as
the product of the Pomeron flux (dependent on $\xpom$ and $t$) and parton
distributions in the Pomeron (dependent on $\beta$ and $Q^{2}$). For $\xpom$ values
substantially larger than 0.01, the contribution of the subleading Reggeon
trajectories may also have to be added.

\subsection{NLO calculation}
\label{sec-mod-nlo}
Predictions for diffractive dijet differential cross sections were
calculated at order $\alpha^{2}_{S}$ with the program {\sc Disent}
\cite{DISENT} adapted for diffractive processes.  The calculations
were performed in the {\sc ${\overline{MS}}$} renormalisation scheme
with five active flavours and the value of the strong coupling
constant set to $\alpha_{s}(M_{Z})=0.118$. The predictions were
obtained with the renormalisation scale, $\mu_{\rm R}$, equal to
$E^{*}_{\rm T,j1}$, where $E^{*}_{\rm T,j1}$ is the
transverse energy of the highest transverse energy jet in the event
(the leading jet) as measured in the $\gamma^{*}p$ centre-of-mass
frame. The factorisation scale was set to $Q^{2}$.

The following dPDFs were used:

\begin{itemize}
\item the ZEUS LPS+charm \cite{LPSfit} - the result of an NLO
  DGLAP QCD fit to the inclusive diffractive structure functions measured
  by the ZEUS experiment with the leading proton spectrometer (LPS).
  In order to better constrain the dPDFs, measurements of $D^{*}$
  production cross section in diffractive DIS \cite{ZEUSDiffrDstar}
  were also included. The fit was restricted to the region
  $\xpom<0.01$;
\item the H1 2006 dPDFs \cite{H1fit2006} - the result of an NLO DGLAP QCD
  fit to a sample of inclusive diffractive structure functions
  measured by the H1 Collaboration. Two different parameterisations
  are available (Fit A and B) which differ in the gluon distribution.
  The fit was restricted to the region $Q^{2}>8.5 \gev^{2},~ \zpom<0.8$.
  Since the H1 measurements were not corrected for the contribution due
  to events where the proton dissociated into a low-mass state, in the
  comparison the calculations were renormalised by a factor $0.87$
  \cite{H1fit2006};
\item the Martin-Ryskin-Watt 2006 (MRW 2006) dPDFs \cite{MRWfit} - 
  the result of a fit to the same data set as for the H1 2006 fit.
  Regge factorisation is assumed only at the input scale.  The dPDFs
  are then evolved with an inhomogeneous evolution equation analogous
  to that for the photon PDFs.  The inhomogenous term 
  accounts for the perturbative Pomeron-to-parton splitting.
\end{itemize}

The only theoretical source of uncertainty considered was that coming
from the NLO calculations. This
uncertainty was estimated by varying $\mu_{\rm R}$ by factors of 0.5
and 2. Uncertainties of more than 20\% were obtained. To compare with
the data, the NLO predictions at the parton level were corrected to the
hadron level using factors extracted from a MC program (see
Section \ref{sec-mod-mc}). The corrections were typically of the order of 10\%.


\section{Monte Carlo simulation}
\label{sec-mod-mc}

Monte Carlo (MC) simulations were used to correct the data for
acceptance and detector effects. Two different MC generators
were used, {\sc Rapgap} \cite{RAPGAP} and {\sc
Satrap} \cite{SATRAP}.

The {\sc Rapgap} MC is based on the factorised-Pomeron approach. The
events were generated using the H1 fit 2 dPDFs \cite{H1fit2}. No
Reggeon contribution was included in this simulation. The
parton-shower simulation is based on the {\sc Meps} \cite{MEPS} model.
Resolved photon processes were also generated using {\sc Rapgap} with
the GRV-G-HO \cite{GRV} photon PDFs. Since the relative contributions
of direct and resolved photon processes to the total cross section are
a priori unknown, the {\sc Rapgap} direct and resolved samples were
weighted in order to best describe the data. The {\sc Rapgap} MC was
also used to extract the hadronisation corrections for the NLO calculation.

{\sc Satrap} is based on the Golec-Biernat--W\uuml sthoff (GBW) dipole
model \cite{SATRAP} and is interfaced to
the {\sc Rapgap} framework.  The parton-shower simulation in {\sc
  Satrap} is based on the Colour Dipole Model (CDM) \cite{CDM}. This
MC does not include the resolved-photon contribution to the
$\gamma^{*}p$ cross section.

To estimate the inclusive DIS background, a sample of events was
generated with {\sc Djangoh} \cite{DJANGOHref}.

All the above MC programs are interfaced to the {\sc
  Heracles} \cite{HERACLESref} event generator for the simulation of
QED radiative processes and to {\sc Jetset} \cite{JETSET} for the
simulation of hadronisation according to the Lund model
\cite{LUNDModel}.  QED radiative corrections were typically between $5$ and $10\%$.

The ZEUS detector response was simulated with a program based on {\sc
  Geant} 3.13 \cite{GEANT}.  The generated events were passed through
the detector simulation, subjected to the same trigger requirements as
the data, and processed by the same reconstruction and offline
programs.  The average of the acceptance-correction values
obtained with {\sc Rapgap} and {\sc Satrap} was used to correct the
data to the hadron level.


\section{Event reconstruction and data selection}
\label{sec-sel}

\subsection{DIS selection}
A three-level trigger system was used to select events online
\cite{ZEUSbluebook,ZEUSTrigger1}.  In the third-level trigger, a DIS
positron candidate and energy deposition in the FPC lower than $20
\gev$ were required. The scattered positron was identified both online
and offline using a neural-network algorithm \cite{SINISTRAalg}. The
reconstruction of the scattered positron variables was
carried out by combining the information from CAL, SRTD and HES.
 In order to select a DIS sample the following requirements were applied \cite{BonatoThesis,TawaraThesis}:
\begin{itemize} 
\item the positron found in the RCAL had to lie outside a rectangular
  area of size [$\mbox{-14, +12}$] cm in $X$ and [$\mbox{-12, +12}$] cm in $Y$, centred
  around the beam pipe. Further cuts on the fiducial area of the
  impact point of the positron on the RCAL surface were applied in
  order to exclude regions with significant inactive material
  \cite{HShapebox};
\item the energy of the scattered positron had to be greater than $10
  \gev$;
\item the vertex of the event had to be in the range
  $\left|Z_{\rm VTX}\right| < 50 \cm$ to reject non-$ep$ background.
\end{itemize}
The four-momentum of the hadronic final-state ${\rm X}$ was reconstructed using
energy-flow objects (EFOs), which combine the information from the CAL
and the CTD\cite{EFO}. The EFOs were corrected for energy losses due
to the inactive material present in the detector \cite{Turcatothesis}. The variable
$\delta=\sum_{i=e,{\rm EFO}}(E_{i}-p_{Z,i})$, where the sum
runs over the scattered positron and all the EFOs, was required to be $45<\delta<65 \gev$.
The variables $E_{i}$ and $p_{Z,i}$ denote the energy and the $Z$-component of
the momentum of each term of the sum.
 
The $Q^{2}$ and $W$ variables were
determined using the double-angle method\cite{DAmethod}. Events
were accepted if $5<Q^{2}<100 \gev^{2}$ and $100<W<250 \gev$.

\subsection{Jet selection}
The $k_{T}$-cluster algorithm in the longitudinal invariant mode
\cite{Ktclusteringjetalg} was applied to the corrected EFOs in the
photon-proton centre-of-mass system ($\gamma^{*}p$ frame) to
reconstruct the jets. The jet variables in the $\gamma^{*}p$ frame are
denoted by a star. After reconstructing the jets, the massless
four-momenta were boosted to the laboratory frame where further energy
corrections were determined and propagated back into the transverse
energy of the jet, $E^{*}_{T,{\rm jet}}$. Such corrections, obtained
from a MC study, improved the correlation between hadron- and
detector-level transverse energy of the jets \cite{BonatoThesis}. The dijet sample was
defined by requiring the events with at least two jets to fulfill the
following constraints:
\begin{itemize}
\item $E^{*}_{T,{\rm j1}}>5 \gev$ and $E^{*}_{T,{\rm j2}}>4 \gev$,
  where the labels ${\rm j1}$ and ${\rm j2}$ refer to the jets
  with the highest and the second highest transverse energy,
  respectively;
\item $-3.5 < \eta^{*}_{\rm jet} < 0$, where $\eta^{*}_{\rm jet}$ is
  the pseudorapidity of any of the jets;
\item the pseudorapidity of the selected jets, boosted to the laboratory frame, had
  to lie in the range $\left|\eta^{\rm LAB}_{\rm jet}\right|<2$.
\end{itemize}

\subsection{Diffractive selection}

Diffractive events are characterised by low values of $\xpom$ and by the
presence of a LRG. The following selection criteria were applied \cite{BonatoThesis,TawaraThesis}:
\begin{itemize}
\item $E_{\rm FPC}<1 \gev$, where $E_{\rm FPC}$ is the total energy in
  the FPC. The requirement of activity compatible with the noise level
  in the angular region covered by the FPC is equivalent to a
  rapidity-gap selection;
\item $\xpomobs < 0.03$ where $\xpomobs$ is the reconstructed value of $\xpom$ and is defined as:
  \begin{displaymath}{\displaystyle \xpomobs=\frac{Q^{2}+M^2_{\rm X}}{Q^{2}+W^{2}}}.\end{displaymath}
  The mass of the diffractive system, $M_{\rm X}$, was reconstructed from
  the EFOs. The cut on $\xpomobs$ reduces the contribution of Reggeon
  exchange and other non-diffractive background.
\end{itemize} 

After these cuts, the selected sample is still contaminated by
diffractive events in which the $p$ dissociated into a low-mass
system. This contamination was estimated by MC studies to be $f_{{\rm
    pdiss}}=(16\pm 4) \%$ \cite{Protondiss} and was subtracted from
the measurements independent of the kinematics.

\begin{figure}
\begin{center}\includegraphics[%
  scale=0.75]{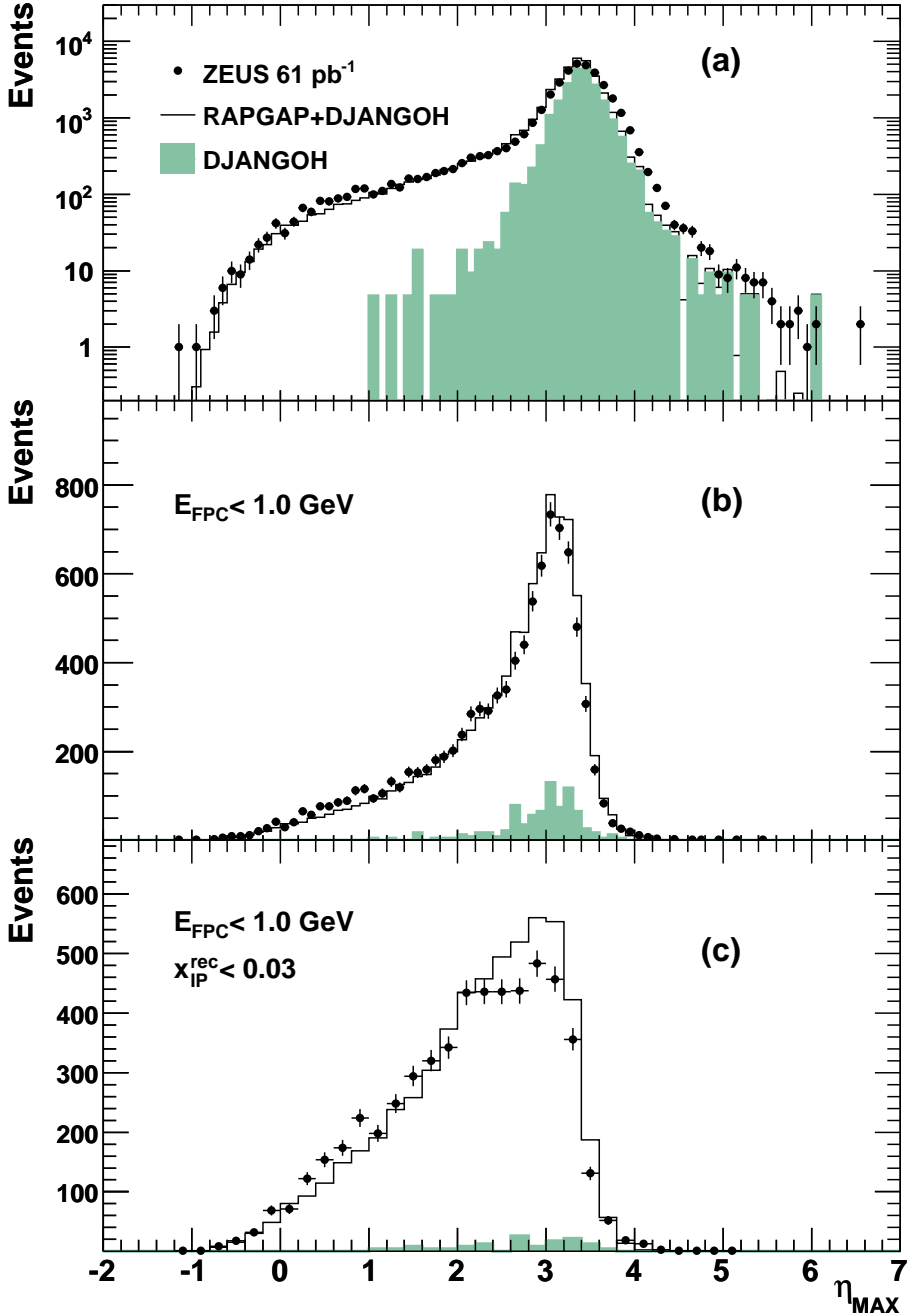}\end{center}

\caption{\label{cap:EtaMaxFig} The measured $\eta_{\rm MAX}$
  distribution (dots) (a) before diffractive selection, (b) after the
  $E_{FPC}$ cut and (c) after adding the $\xpomobs$ cut. Also shown
  are area-normalised MC expectations obtained by fitting the relative
  amount of {\sc Rapgap} and {\sc Djangoh} to give the best
  description of the data before any diffractive selection.}
\end{figure}

The contamination of the non-diffractive background as a function of
the applied diffractive selection cuts is shown in
Fig.~\ref{cap:EtaMaxFig}, through the distribution of $\eta_{\rm
  MAX}$, where $\eta_{\rm MAX}$ is the pseudorapidity in the
laboratory frame of the most forward EFO with energy higher than $400
\mev$, before and after applying cuts on the $E_{\rm FPC}$ and on
$\xpomobs$. The disagreement between the measured and the simulated
distributions is the reason for not applying any requirement on
$\eta_{\rm MAX}$, as was done in previous analyses
\cite{ZEUSincldiffDIS,ZEUSDiffrDstar,ZEUSDstarPHP}. After the
$E_{\rm FPC}$ and $\xpomobs$ cuts, the non-diffractive background
from {\sc Djangoh} was estimated to be 2.4\% of the total
selected events and neglected in further analysis. After all cuts,
5539 events remained.



\section{Systematic uncertainties}
\label{sec-syst}

The systematic uncertainties of the measured cross sections were
calculated by varying the cuts and the analysis procedure. The
systematic checks were the following:

\begin{itemize}

\item the energy measured by the CAL was varied by $\pm 3\%$ in
  the MC to take into account the uncertainty on the CAL
  calibration, giving one of the largest uncertainties.
  Deviations from nominal cross section values were of the order of
  $\pm 5\%$, but reached $\sim 15\%$ in some bins;
\item the energy scale of the scattered positron was varied in the MC
  by its uncertainty, $\pm 2\%$. The resulting variation of the cross
  sections was always below $\pm 3\%$;
\item the position of the SRTD was changed in the MC by $\pm 2 \mm$ in
  all directions to account for the uncertainty on its alignment. The
  change along the $Z$ direction gave the largest effect and in a few
  bins caused a cross section variation of $\pm 2\%$;
\item the model dependence of the acceptance corrections was estimated
  by using separately {\sc Rapgap} and {\sc Satrap} for unfolding the
  data. The variations from the central value (obtained using the
  average between {\sc Rapgap} and {\sc Satrap}) were typically of the
  order of $\pm 5\%$ but reached $\sim \pm 10\%$ in some bins.
\end{itemize}

The above systematic uncertainties, except those related to the energy
scale of the calorimeter, were added in quadrature to determine the
total systematic uncertainty. The uncertainties due to the energy
scale and the proton dissociation subtraction ($\pm 4\%$) were added
in quadrature and treated as correlated systematics. The energy scale
uncertainty is quoted separately in the tables.

The stability of the measurement was checked by varying the
selection cuts as follows:
\begin{itemize}
\item the cut on the FPC energy was varied by $\pm 100 \mev$ in the
  MC;
\item the cut on the scattered-positron energy was lowered from $10$
  to $8 \gev$;
\item the fiducial region for the positron selection was
  enlarged and reduced by $0.5~{\rm cm}$;
\item the lower cut on $\delta$ was changed from $45$ to $43
  \gev$.
\end{itemize}

The variations of the cross section induced by these stability checks
were small, within $\pm 2\%$, and were added in quadrature to the
total systematic uncertainty. The uncertainty on the luminosity
measurement ($2.25\%$) was not included.

The measurement was repeated with the addition of a cut on the value of
$\eta_{\rm MAX}$. This estimates the uncertainty on the purity of the
diffractive selection. A cut of $\eta_{\rm MAX}<2.8$ was applied.  The
cross sections increased by $\sim 5\%$ and the change was concentrated
at high values of $\xpomobs$. No significant dependence on other
variables was observed. This variation is listed in the tables for
completeness but not included in the quoted uncertainties of the measurement.



\section{Results and discussion}
\label{sec-dis}
The single- and double-differential cross sections for the production
of dijets in diffractive DIS have been measured for $5 < Q^{2}<100$
\gev$^{2}$, $100< W<250 \gev$ and $\xpom<0.03$, for jets in the
pseudorapidity region $-3.5< \eta^{*}_{\rm jet} <0$, with
$E^{*}_{T,\rm j1}>5 \gev$ and $E^{*}_{T,\rm j2}>4\gev$.  The cross
sections refer to jets of hadrons and are corrected for QED effects.

The measured total cross section (given in Table~\ref{xsec_tabTOT}) is:
\begin{center}\begin{displaymath}
\sigma(ep\rightarrow ep+{\rm j1 + j2 + X'})= 89.7 \pm 1.2 {\rm (stat)} \;^{+3.2}_{-5.3}{\rm (syst.)} \;^{+5.1}_{-3.7}{\rm (corr.)}~{\rm pb}.
\end{displaymath}\end{center}
The values of the differential cross sections are averaged over the bin in which they
are presented. For any variable $\kappa$, the cross section was
determined as
\begin{equation}
\frac{d\sigma}{d\kappa}=C\, \frac{N_{D}(1-f_{{\rm pdiss}})}{\Lumi \,\Delta\kappa},
\end{equation}

where $N_{D}$ is the number of data events in a bin, $C$ includes the
effects of the acceptance and the QED correction factor as determined
from MC, $\Lumi$ is the integrated luminosity and $\Delta\kappa$ is
the bin width.

The differential cross sections were measured as a function of
$Q^{2}$, $W$, $\xpomobs$, $\beta$, $M_{\rm X}$, \etjj, \etajj,
$\zpomobs$ and $\xgammaobs$. The variable $\etjj$ ($\etajj$) stands
for both $E_{\rm T,j1}^{*}$ ($\eta^{*}_{\rm j1}$) and $E_{\rm
  T,j2}^{*}$ ($\eta^{*}_{\rm j2}$) - in the corresponding cross
section, it thus contributes two entries per event. The variable
$\zpomobs$ is an estimator of $\zpom$ and is calculated as

\begin{figure}
\begin{center}\includegraphics[%
  width=1.0\linewidth]{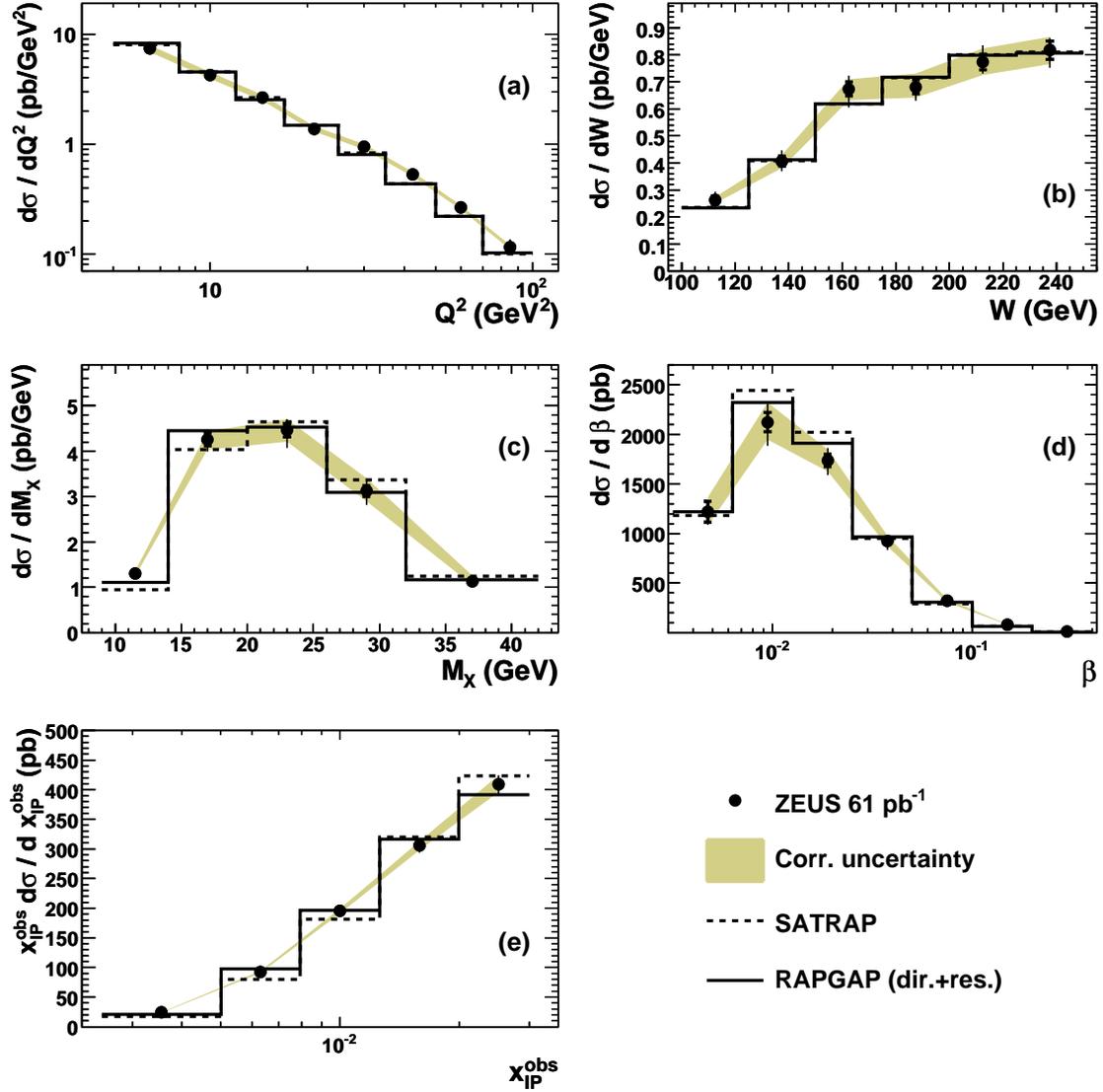}\end{center}

\caption{\label{cap:LO_fig1} Measured differential cross section
  (dots) as a function of (a) $Q^{2}$, (b) $W$, (c) $M_{X}$, (d)
  $\beta$ and (e) $\xpomobs$. The inner error bars represent the
  statistical uncertainty and the outer error bars represent the
  statistical and uncorrelated systematic uncertainties added in
  quadrature. The shaded band represents the correlated uncertainty.
  For comparison the area-normalised predictions of the {\sc Rapgap}
  (solid lines) and the {\sc Satrap} (dashed lines) MC models are also
  shown.}
\end{figure}

\begin{center}\begin{displaymath}\zpomobs={\displaystyle
      \frac{Q^{2}+M_{\rm jj}^{2}}{Q^{2}+M_{\rm X}^{2}}},\end{displaymath}\end{center}

where $M_{\rm jj}$ is the invariant mass of the dijet system. The
estimator of $\xgamma$, $\xgammaobs$, is
\begin{center}\begin{displaymath}\xgammaobs=\frac{ {\displaystyle E^{\rm LAB}_{\rm T,j1}e^{-\eta^{\rm LAB}_{\rm j1}} + E^{\rm LAB}_{\rm T,j2}e^{-\eta^{\rm LAB}_{\rm j2}} }}{{\displaystyle \sum_{\rm hadr}(E_{i}-p_{Z,i})}},\end{displaymath}\end{center}

where the sum in the denominator runs over all the hadrons. The values
of the differential cross sections are presented in
Tables~\ref{xsec_tabQ2}-\ref{xsec_tabzpomDD} and shown in Figs.~\ref{cap:LO_fig1} and \ref{cap:LO_fig2}.

\subsection{Comparison to Monte Carlo models}

\begin{figure}

\begin{center}\includegraphics[%
  width=1.0\linewidth]{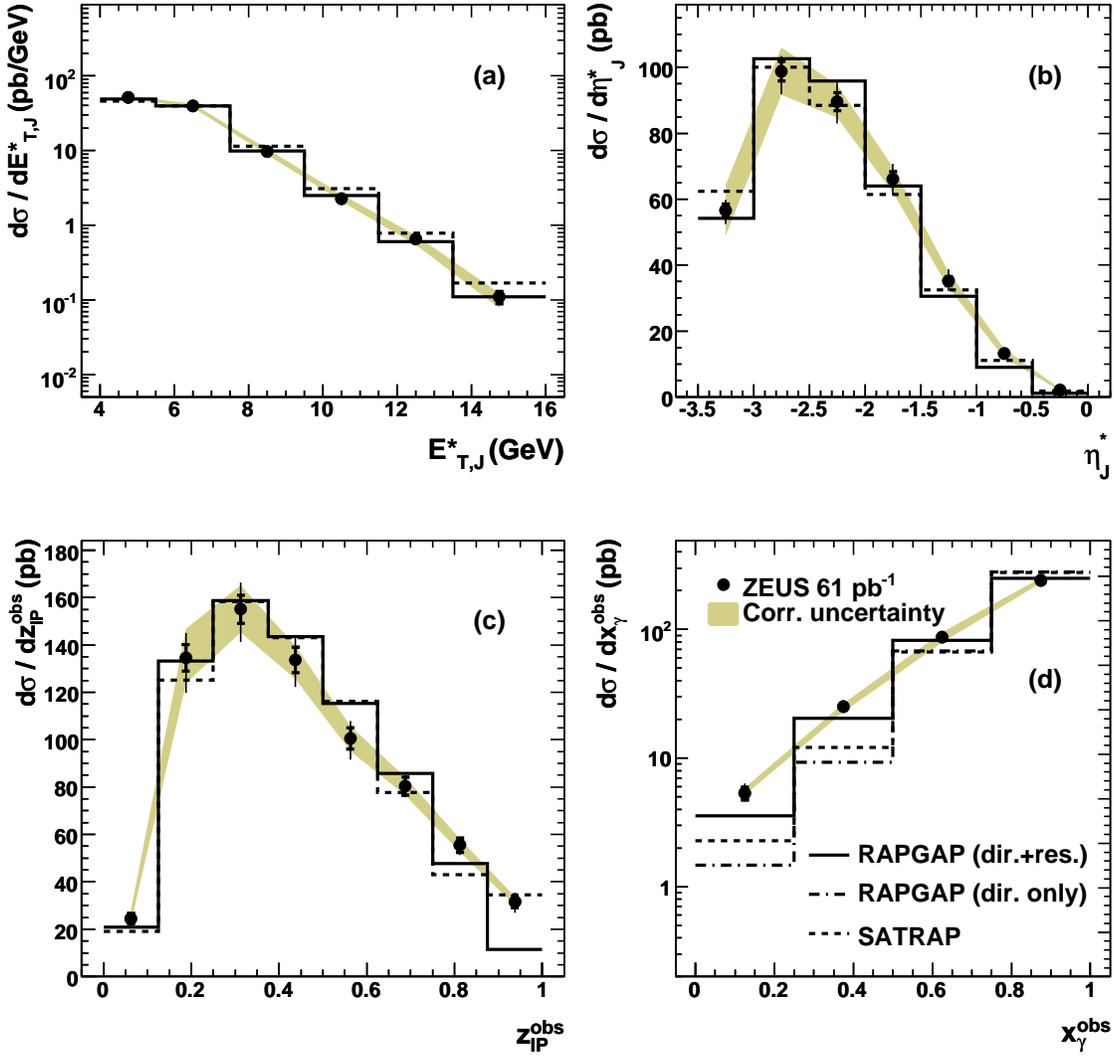}\end{center}

\caption{\label{cap:LO_fig2}Measured differential cross section as a function of
  (a) \etjj , (b) \etajj , (c) $\zpomobs$ and (d) $\xgammaobs$. The
  dashed-dotted line represents the area-normalised {\sc Rapgap} with
  only the direct photon contribution. Other details as in the caption of
  Fig.~\ref{cap:LO_fig1}.}

\end{figure}

The {\sc Rapgap} and {\sc Satrap} MC programs are compared to the
measured cross sections in Figs.~\ref{cap:LO_fig1} and
\ref{cap:LO_fig2}.  Since the MC predictions are not expected to
describe the normalisation, the cross sections predicted by both MCs
were normalised to the data.  The total correlated uncertainty is
shown as a shaded band in the figures.  The comparison with MC
predictions shows in general a reasonably good agreement with the
shape of the data. The $\etjj$ distribution is a steeply falling
function as expected in pQCD (Fig.~\ref{cap:LO_fig2}a) and the jets
tend to populate the $\gamma^{*}$ fragmentation region.

The most prominent features of the data are the rise of the cross
section with $\xpomobs$, the peak at $\zpomobs$ $\sim 0.3$ and the tail of
the cross section at low $\xgammaobs$ values.  The requirement of two
jets with high $E_{\rm T}$ suppresses the contribution of low values of
$\xpomobs$. The relatively low value of the peak position in the
$\zpomobs$ distribution indicates that in the majority of the events the
dijet system is accompanied by additional hadronic activity. A
disagreement between data and {\sc Rapgap} is observed at high
$\zpomobs$. In the high $\zpomobs$ region, {\sc Rapgap} underestimates the
number of events while {\sc Satrap} agrees with the data, possibly
because of the presence of a mechanism for exclusive direct
production. Most of the events are produced at large $\xgammaobs$ as
expected in DIS. At low $\xgammaobs$, the description by
{\sc Rapgap} is improved by the addition of the resolved photon
contribution (16\%).

\subsection{Comparison to NLO QCD predictions}

In Table~\ref{xsec_tabTOT}, the four NLO predictions described in
Section~\ref{sec-mod-nlo} are compared to the measured total cross
section. The central values of the predictions using the $\honefitB$
and MRW 2006 dPDFs give the best description, while those using the
$\honefitA$ and the ZEUS LPS+charm dPDFs are higher in
normalisation.

\begin{figure}
\begin{center}\includegraphics[%
  scale=0.75]{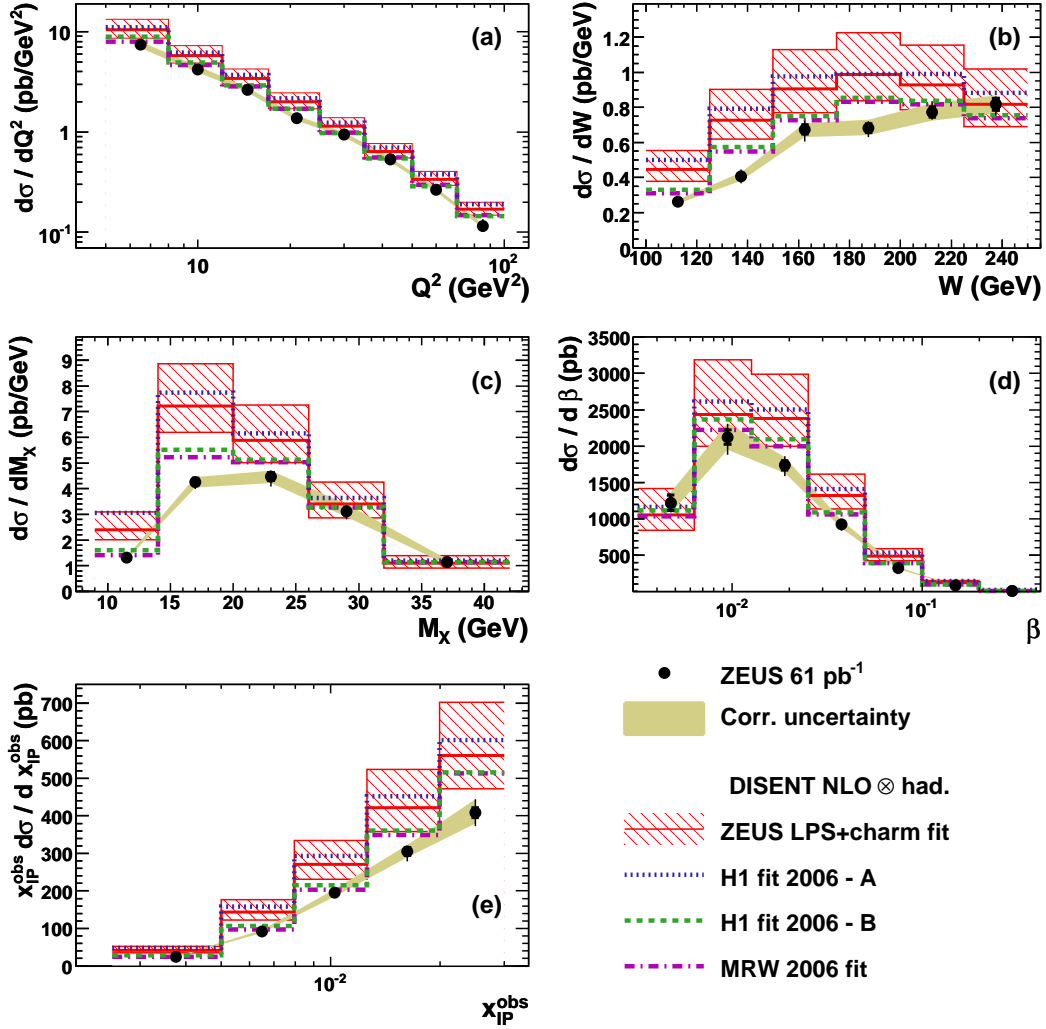}\end{center}

\caption{\label{cap:NLO_fig1}Measured differential cross section as a
  function of (a) $Q^{2}$, (b) $W$, (c) $M_{X}$, (d) $\beta$ and (e)
  $\xpomobs$ compared to the NLO predictions obtained using the
  available dPDFs, as indicated in the figure.  The hatched area
  indicates the theoretical uncertainty of the predictions estimated
  using the ZEUS LPS+charm dPDFs. Other details as in the caption of
  Fig.~\ref{cap:LO_fig1}.}
\end{figure}

\begin{figure}
\begin{center}\includegraphics[%
  scale=0.75]{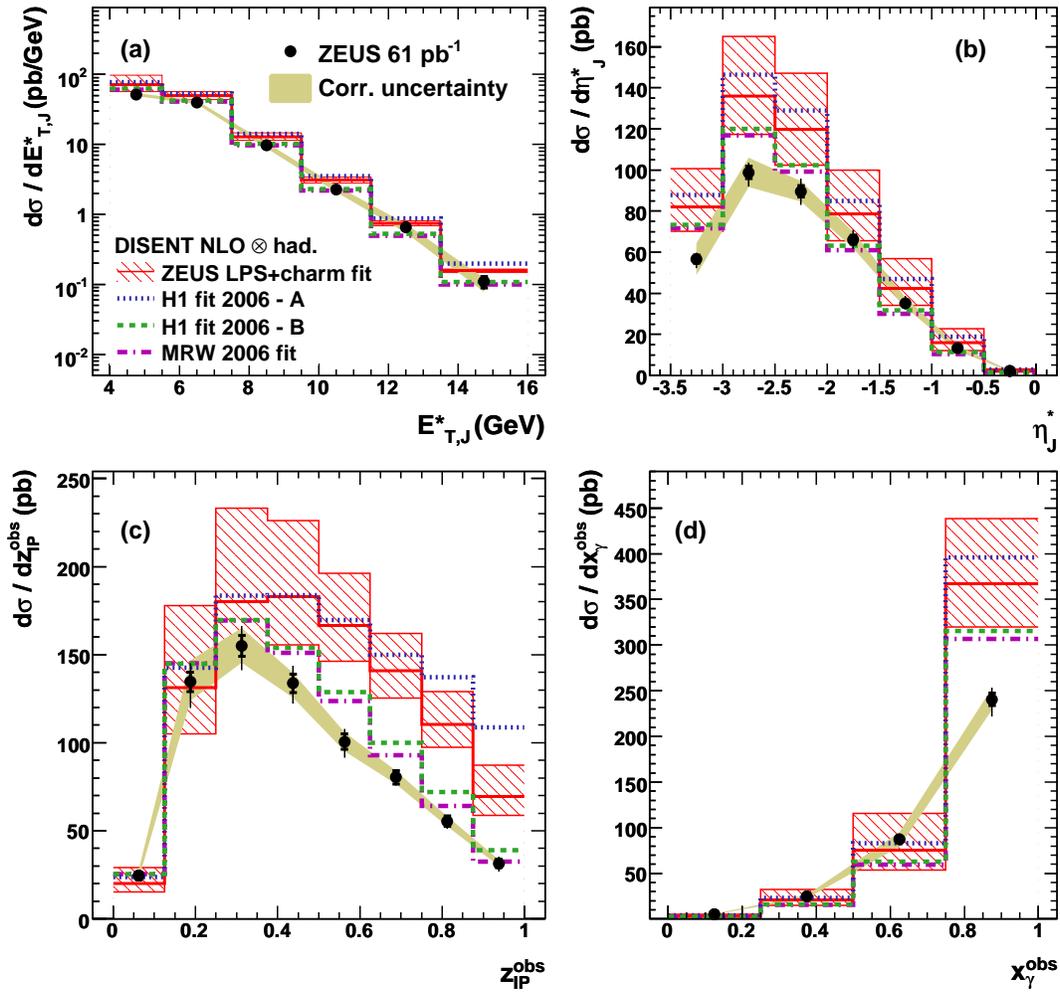}\end{center}

\caption{\label{cap:NLO_fig2}Measured differential cross section as a function of
  (a) \etjj, (b) \etajj , (c) $\zpomobs$ and (d) \xgammaobs compared to the
  NLO prediction obtained using the available dPDFs.  Other details as in the caption of
  Fig.~\ref{cap:NLO_fig1}.}
\end{figure}

\begin{figure}
\begin{center}\includegraphics[%
  width=1.0\linewidth]{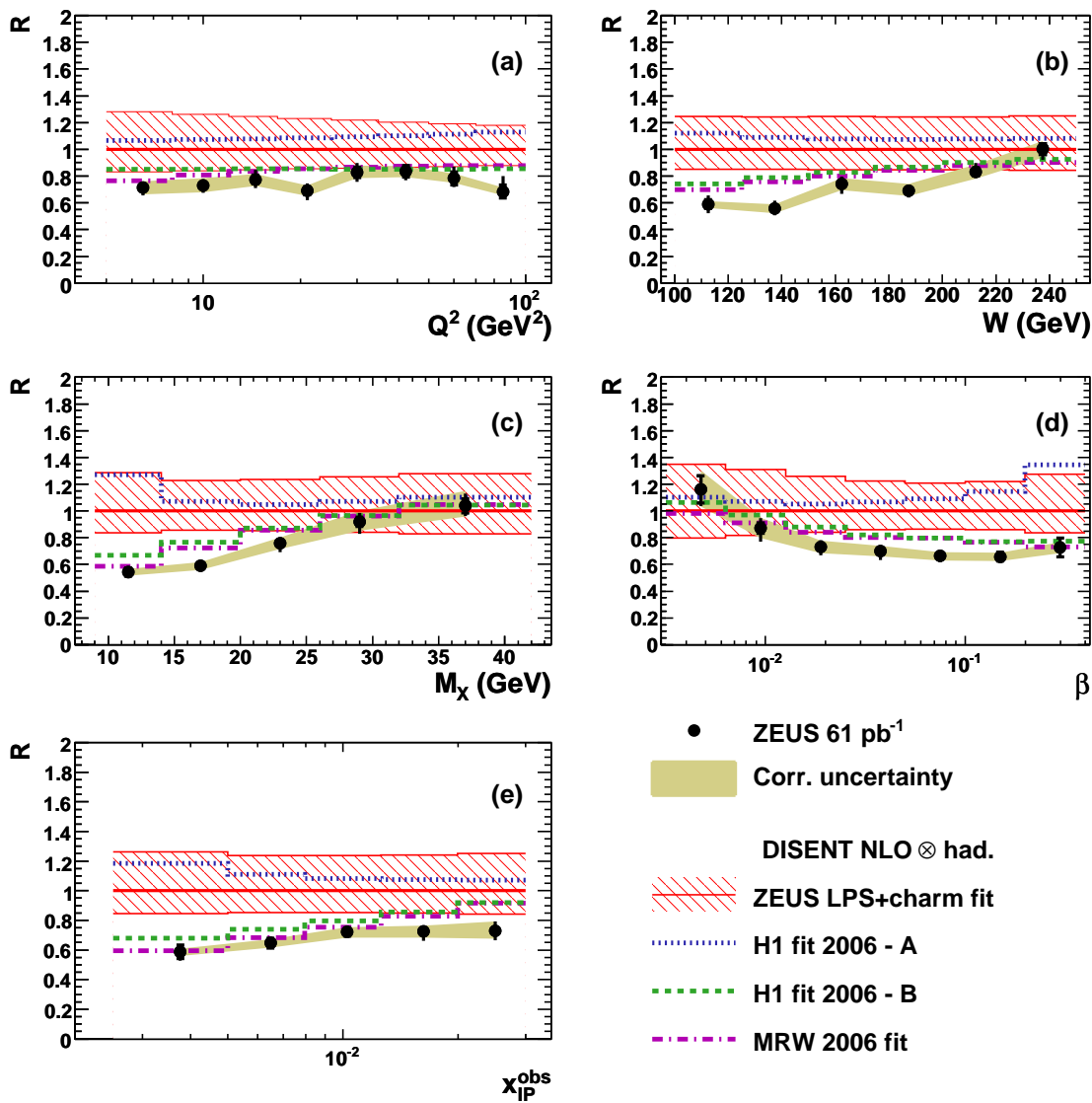}\end{center}

\caption{\label{cap:NLO_ratio1} Ratio, R, of the data to the NLO
  prediction using the ZEUS LPS+charm dPDFs (dots) as function of (a)
  $Q^{2}$, (b) $W$, (c) $M_{X}$, (d) $\beta$ and (e) $\xpomobs$. Also
  shown is the ratio of NLO calculations with other dPDFs to ZEUS
  LPS+charm. Other details as in the caption of
  Fig.~\ref{cap:NLO_fig1}.}

\end{figure}

\begin{figure}
\begin{center}\includegraphics[%
  width=1.0\linewidth]{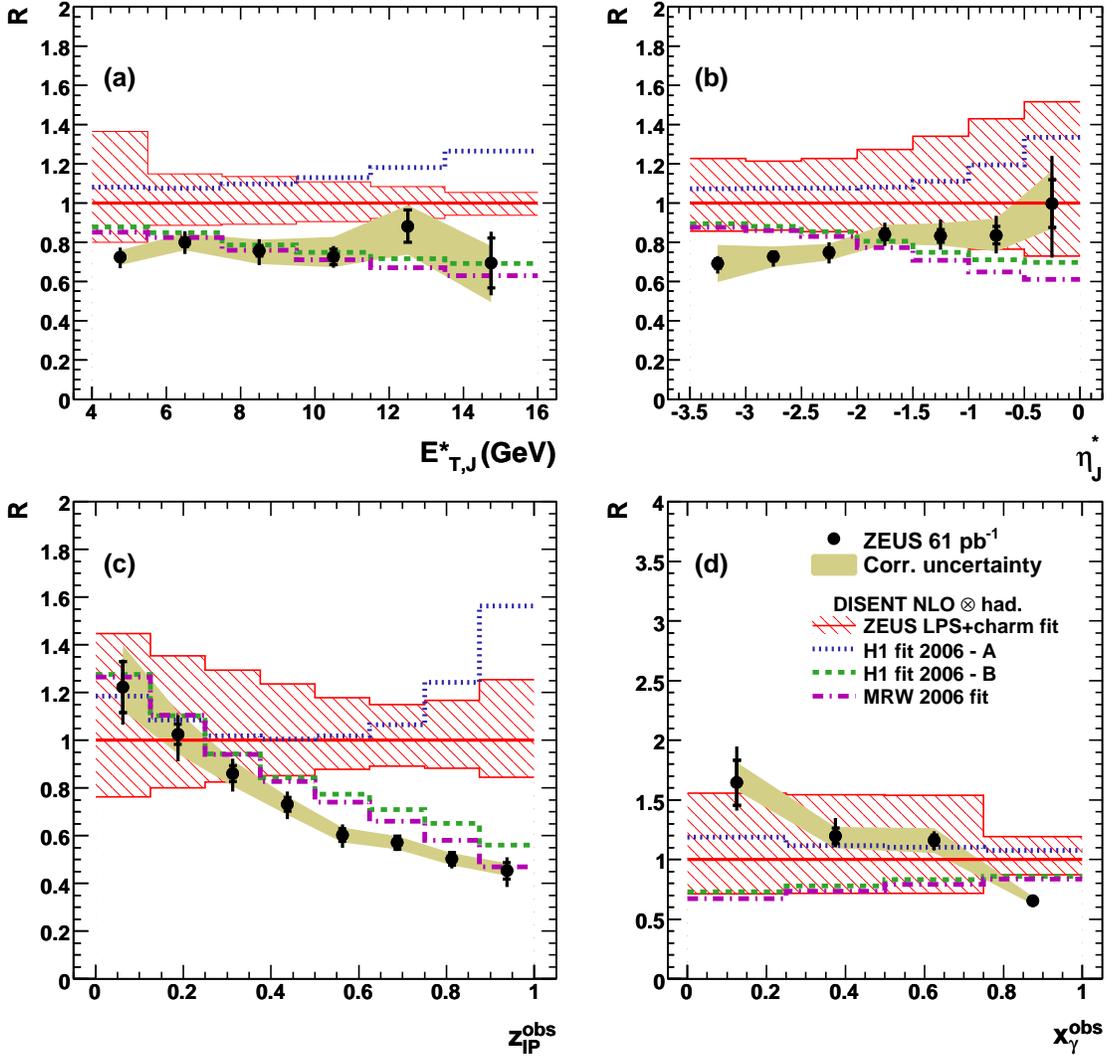}\end{center}

\caption{\label{cap:NLO_ratio2}Ratio, R, of the data to the NLO
  prediction using the ZEUS LPS+charm dPDFs (dots) as function of (a)
  \etjj, (b) \etajj, (c) $\zpomobs$ and (d) \xgammaobs. Other details as in the caption of
  Fig.~\ref{cap:NLO_ratio1}.}
\end{figure}

The NLO predictions for the differential cross section are compared to
the data in Figs.~\ref{cap:NLO_fig1} and \ref{cap:NLO_fig2}. The
estimated theoretical uncertainties are shown only for the
calculations using the ZEUS LPS+charm dPDFs and are similar for all
the other calculations. For ease of comparison the ratios of data to
the ZEUS LPS+charm prediction are presented in
Figs.~\ref{cap:NLO_ratio1} and \ref{cap:NLO_ratio2}. The variation due
to the choice of the dPDFs is displayed with respect to the ZEUS
LPS+charm in the same figure. In general the shape of the measured
cross section is described by the NLO calculations within the
theoretical uncertainties. However, only the predictions using the
$\honefitB$ and MRW 2006 dPDFs are able to describe satisfactorily the
data over the entire kinematic range.

\begin{figure}
\begin{center}\includegraphics[%
  width=1.0\linewidth]{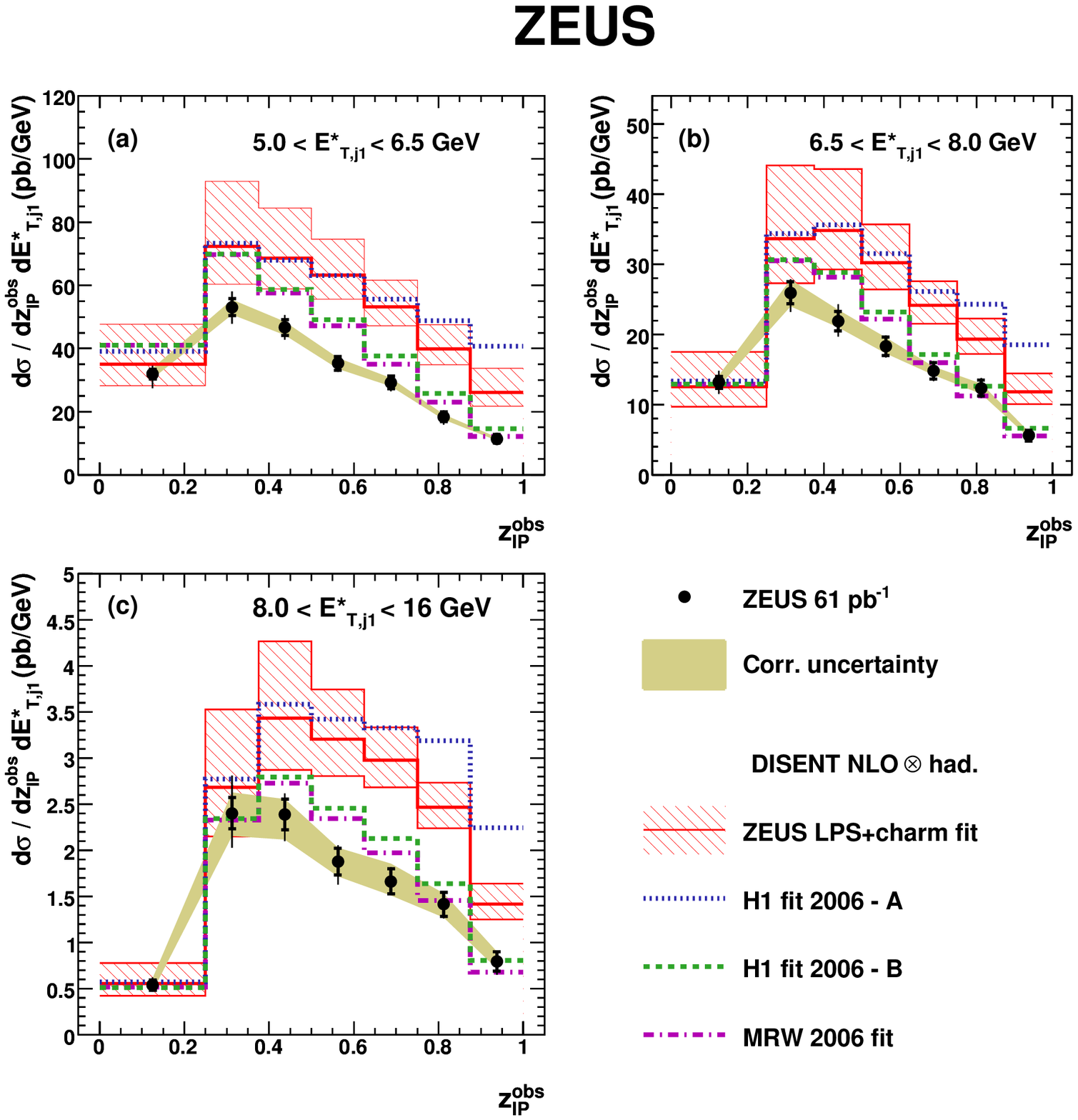}\end{center}

\caption{\label{cap:NLO_DD_1}Measured differential cross section
  as a function of $\zpomobs$ in different regions of
  $\ensuremath{E_{T,j1}^{*}}$ (dots). Other details as in the caption of
  Fig.~\ref{cap:NLO_fig1}.}
\end{figure}

\begin{figure}
\begin{center}\includegraphics[%
  width=1.0\linewidth]{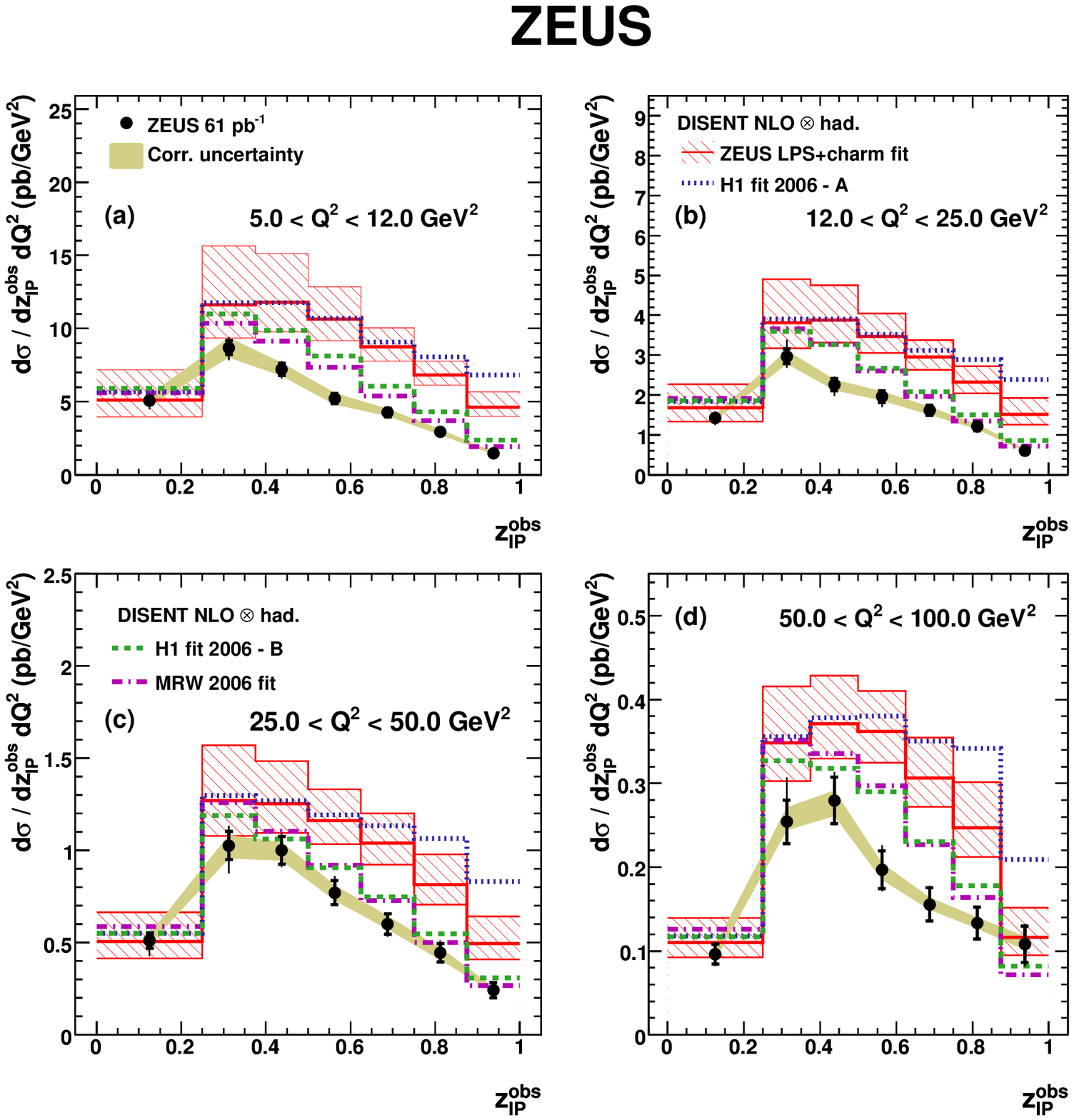}\end{center}

\caption{\label{cap:NLO_DD_2}Measured differential cross section
  as a function of $\zpomobs$ in different regions of
  $Q^{2}$ (dots). Other details as in the caption of
  Fig.~\ref{cap:NLO_fig1}.}
\end{figure}

The NLO predictions for the differential cross section are compared to
the data in Figs.~\ref{cap:NLO_DD_1} and \ref{cap:NLO_DD_2}, where the
$\zpomobs$ distribution is shown for different regions of $E_{\rm
  T,j1}^{*}$ and $Q^{2}$. Within the theoretical
uncertainties, the $\honefitB$ and MRW 2006 dPDFs are compatible with
the data. Since the major difference between the $\honefitB$ and Fit A
is in the gluon dPDF, these data have a significant potential
to further constrain the gluon dPDF.

\section{Conclusions}
The single- and double-differential cross sections for the production
of dijets in diffractive DIS have been measured with the ZEUS detector
in the kinematic region $5 < Q^{2}<100$ \gev$^{2}$, $100< W<250 \gev$
and $\xpom<0.03$, requiring at least two jets with $E^{*}_{T,\rm
  jet}>4\gev$ in the pseudorapidity region $-3.5< \eta^{*}_{\rm jet}
<0.0$ and the highest $E^{*}_{T}$ jet with $E^{*}_{T,\rm j1}>5 \gev$.

Two leading-logarithm parton-shower models, {\sc Rapgap} and {\sc
  Satrap}, describe the shape of the measured cross sections well. The
measured cross sections are able to discriminate between NLO QCD
calculations based on different dPDFs, showing a satisfactory agreement
with the calculations using the $\honefitB$ and MRW 2006 dPDFs.
This lends further support to the validity of QCD factorisation in
hard diffractive scattering. Since the dPDFs used differ mostly in the
gluon content, these data may have a significant potential to
constrain the diffractive gluon distribution.

\section{Acknowledgments}
We are grateful to the DESY Directorate for their strong support and
encouragement. The effort of the HERA machine group is gratefully
acknowledged. We thank the DESY computing and network services for
their support. The design, construction and installation of the ZEUS
detector have been made possible by the efforts of many people not
listed as authors. We thank G.~Watt and H.~Jung for valuable
discussions and suggestions.



\begin{table}[p]
\begin{center}
\begin {tabular}{|c|c|c|c|c|c|c|}
\hline
$\;$ & $\sigma$ &  $\delta_{{\rm stat}}$ & $\delta_{{\rm syst}}$ & $\delta_{{\rm ES}}$ & $\delta_{{\rm theor}}$&$\Delta_{{\rm DIFFR}}$\\
$\;$ & (pb) &  (pb) & (pb) & (pb) & (pb)& (pb)\\
\hline  \hline
Data  & 89.7 & 1.2& $\;^{+3.2}_{-5.3}$ & $\;^{+5.1}_{-3.7}$ & -- &+4.0 \\
ZEUS LPS+charm & 120.3 & --& -- & -- & $\;^{+29.4}_{-18.3}$ & --\\
\honefitA  & 130.2 & --& -- & -- & $\;^{+31.2}_{-19.9}$ & --\\
\honefitB  & 102.5 & --& -- & -- & $\;^{+24.7}_{-15.6}$ & --\\
MRW 2006  & 99.3 & --& -- & -- & $\;^{+23.4}_{-14.7}$ & --\\
\hline
\hline
\end{tabular}
\vspace*{1cm}
\caption{\label{xsec_tabTOT} Total cross section for the production of diffractive dijets compared to expectations of NLO calculations using various dPDFs as indicated in the Table. The cross section is given for jets with  $E^{*}_{T,\rm j1}>5 \gev$, $E^{*}_{T,\rm j2}>4 \gev$, $-3.5< \eta^{*}_{\rm jet} <0$ and in the range of $5 < Q^{2}<100 \gev^{2}$, $100< W<250 \gev$ and $\xpom<0.03$. The statistical, $\delta_{{\rm stat}}$, uncorrelated systematic, $\delta_{{\rm syst}}$, and energy scale uncertainties, $\delta_{{\rm ES}}$, are quoted separately. The theoretical uncertainty on the NLO calculations, $\delta_{{\rm theor}}$, is quoted in the sixth column. The difference with the measured cross section with and without $\eta_{\rm MAX}$ cut, $\Delta_{{\rm DIFFR}}$, is presented in the last column. The uncertainties on the proton dissociation subtraction and the luminosity measurement are not presented in the table.}
\vspace*{1cm}
\end{center}
\end{table}

\begin{table}[p]
\begin{center}
\begin {tabular}{|r@{,$\;$}l|r@{.}l|l r@{.}l|p{2.0cm}|p{2.0cm}|r@{.}l|}
\hline
\multicolumn{2}{|c|}{$Q^{2}$ bin} &  \multicolumn{2}{|c|}{$d\sigma/dQ^{2}$}  &  \multicolumn{3}{|c|}{$\delta_{{\rm stat}}$} & $\delta_{{\rm syst}}$ & $\delta_{{\rm ES}}$ &  \multicolumn{2}{|c|}{$\Delta_{{\rm DIFFR}}$}\\
\multicolumn{2}{|c|}{$(\gev^{2})$}   &  \multicolumn{2}{|c|}{(pb/$\gev^{2}$)}   &  \multicolumn{3}{|c|}{(pb/$\gev^{2}$)}& (pb/$\gev^{2}$)& (pb/$\gev^{2}$)&  \multicolumn{2}{|c|}{(pb/$\gev^{2}$)}\\
\hline 
\hline
5&8 & 7&4 & $\pm$&0&3 & $\;^{+0.3}_{-0.5}$ &  $\;^{+0.5}_{-0.5}$ & 0&1\\
\hline
8&12 & 4&2 & $\pm$& 0&2 & $\;^{+0.2}_{-0.3}$ &  $\;^{+0.3}_{-0.3}$ & 0&1\\
\hline
12&17 & 2&6 & $\pm$& 0&1 & $\;^{+0.2}_{-0.2}$ &  $\;^{+0.2}_{-0.2}$ & 0&2\\
\hline
17&25 & 1&38 & $\pm$& 0&06 & $\;^{+0.09}_{-0.13}$ &  $\;^{+0.08}_{-0.08}$ & 0&06\\
\hline
25&35 & 0&94 & $\pm$& 0&04 & $\;^{+0.07}_{-0.07}$ &  $\;^{+0.06}_{-0.05}$ & 0&06\\
\hline
35&50 & 0&53 & $\pm$& 0&03 & $\;^{+0.02}_{-0.03}$ &  $\;^{+0.03}_{-0.03}$ & 0&01\\
\hline
50&70 & 0&27 & $\pm$& 0&02 & $\;^{+0.02}_{-0.01}$ &  $\;^{+0.01}_{-0.01}$ & 0&02\\
\hline
70&100 & 0&116 & $\pm$& 0&008 & $\;^{+0.018}_{-0.003}$ &  $\;^{+0.005}_{-0.005}$ & 0&018\\
\hline
\hline
\end{tabular}
\vspace*{1cm}
\caption{\label{xsec_tabQ2}
  Values of the differential cross section as a function of
  $Q^{2}$ for the production of diffractive dijets. The range over which the cross section is averaged is given in the first column.
  Other details as in the caption of Table \ref{xsec_tabTOT}.}

\vspace*{1cm}
\begin {tabular}{|r@{,$\;$}l|r@{.}l|l r@{.}l|p{2.0cm}|p{2.0cm}|r@{.}l|}
\hline
\multicolumn{2}{|c|}{$W$ bin} &  \multicolumn{2}{|c|}{$d\sigma/dW$} &  \multicolumn{3}{|c|}{$\delta_{{\rm stat}}$} & $\delta_{{\rm syst}}$ & $\delta_{{\rm ES}}$&  \multicolumn{2}{|c|}{$\Delta_{{\rm DIFFR}}$}\\
\multicolumn{2}{|c|}{$(\gev)$} &  \multicolumn{2}{|c|}{(pb/\gev)}  &  \multicolumn{3}{|c|}{(pb/\gev)} & (pb/\gev) & (pb/\gev) &  \multicolumn{2}{|c|}{(pb/\gev)}\\
\hline  \hline
100 & 125 & 0&26 & $\pm$&0&01 & $\;^{+0.03}_{-0.03}$ &  $\;^{+0.01}_{-0.01}$ & 0&01\\
\hline
125 & 150 & 0&41 & $\pm$&0&02 & $\;^{+0.04}_{-0.03}$ &  $\;^{+0.02}_{-0.03}$ & 0&03\\
\hline
150 & 175 & 0&67 & $\pm$&0&03 & $\;^{+0.04}_{-0.06}$ &  $\;^{+0.04}_{-0.04}$ & 0&03\\
\hline
175 & 200 & 0&68 & $\pm$&0&02 & $\;^{+0.03}_{-0.04}$ &  $\;^{+0.05}_{-0.04}$ & 0&01\\
\hline
200 & 225 & 0&77 & $\pm$&0&03 & $\;^{+0.06}_{-0.03}$ &  $\;^{+0.05}_{-0.05}$ & 0&05\\
\hline
225 & 250 & 0&82 & $\pm$&0&03 & $\;^{+0.03}_{-0.06}$ &  $\;^{+0.05}_{-0.05}$ & 0&02\\
\hline
\hline
\end{tabular}
\vspace*{1cm}
\caption{\label{xsec_tabW}Values of the differential cross section as a function of
  $W$. Other details as in the caption of Table~\ref{xsec_tabQ2}.}
\vspace*{1cm}
\end{center}
\end{table}

\begin{table}[p]
\begin{center}
\begin {tabular}{|r@{,$\;$}l|r@{.}l|l r@{.}l|p{2.0cm}|p{2.0cm}|r@{.}l|}
\hline
\multicolumn{2}{|c|}{$M_{X}$ bin} &\multicolumn{2}{|c|}{ $d\sigma/dM_{X}$}  & \multicolumn{3}{|c|}{$\delta_{{\rm stat}}$} & $\delta_{{\rm syst}}$ & $\delta_{{\rm ES}}$ & \multicolumn{2}{|c|}{$\Delta_{{\rm DIFFR}}$}\\
\multicolumn{2}{|c|}{$(\gev)$} &\multicolumn{2}{|c|}{ (pb/$\gev$)}  & \multicolumn{3}{|c|}{ (pb/$\gev$)} & (pb/$\gev$) & (pb/$\gev$) & \multicolumn{2}{|c|}{ (pb/$\gev$)}\\
\hline  \hline
9 & 14 & 1&31 & $\pm$& 0&07 & $\;^{+0.02}_{-0.08}$ &  $\;^{+0.05}_{-0.06}$ & -0&03\\
\hline
14 & 20 & 4&3 & $\pm$& 0&1 & $\;^{+0.2}_{-0.2}$ &  $\;^{+0.2}_{-0.2}$ & 0&1\\
\hline
20 & 26 & 4&5 & $\pm$& 0&1 & $\;^{+0.2}_{-0.4}$ &  $\;^{+0.2}_{-0.2}$ & 0&0\\
\hline
26 & 32 & 3&1 & $\pm$& 0&1 & $\;^{+0.2}_{-0.3}$ &  $\;^{+0.3}_{-0.2}$ & -0&1\\
\hline
32 & 42 & 1&13 & $\pm$& 0&05 & $\;^{+0.08}_{-0.06}$ &  $\;^{+0.12}_{-0.09}$ & 0&07\\
\hline
\hline
\end{tabular}
\vspace*{1cm}
\caption{\label{xsec_tabmx}Values of the differential cross sections with respect to
  $M_{X}$. Other details as in the caption of Table~\ref{xsec_tabQ2}. }
\vspace*{1cm}

\begin {tabular}{|r@{,$\;$}l|r@{}l|l r@{}l|p{2.0cm}|p{2.0cm}|r@{}l|}
\hline
\multicolumn{2}{|c|}{$\beta$ bin} &\multicolumn{2}{|c|}{$d\sigma/d\beta$}  &  \multicolumn{3}{|c|}{$\delta_{{\rm stat}}$} & $\delta_{{\rm syst}}$ & $\delta_{{\rm ES}}$ & \multicolumn{2}{|c|}{$\Delta_{{\rm DIFFR}}$} \\
\multicolumn{2}{|c|}{($\times\,10^{-2}$)} &\multicolumn{2}{|c|}{(pb)}       &  \multicolumn{3}{|c|}{(pb)} & (pb) & (pb) & \multicolumn{2}{|c|}{(pb)}\\
\hline  \hline
0.32 & 0.63 & 1220$\;$& & $\pm$& 102$\;$& & $\;^{+30}_{-75}$ &  $\;^{+148}_{-135}$ & -69$\;$&\\
\hline
0.63 & 1.26 & 2124$\,$& & $\pm$&  94$\;$& & $\;^{+153}_{-221}$ &  $\;^{+196}_{-177}$ & -11$\;$&\\
\hline
1.26 & 2.51 & 1736$\,$& & $\pm$&  62$\;$& & $\;^{+108}_{-133}$ &  $\;^{+112}_{-109}$ & 46$\;$&\\
\hline
2.51 & 5.01 & 923$\,$& & $\pm$&  32$\;$& & $\;^{+40}_{-83}$ &  $\;^{+55}_{-50}$ & 3$\;$&\\
\hline
5.01 & 10.00 & 324$\;$& & $\pm$&  12$\;$& & $\;^{+9}_{-18}$ &  $\;^{+14}_{-17}$ & 3$\;$&\\
\hline
10.00 & 19.95 & 81.&8 & $\pm$& 3.&8 & $\;^{+4.3}_{-2.7}$ &  $\;^{+3.5}_{-4.1}$ & 4.&1\\
\hline
19.95 & 39.81 & 9.&7 & $\pm$& 0.&8 & $\;^{+0.5}_{-0.5}$ &  $\;^{+0.5}_{-0.6}$ & 0.&4\\

\hline
\hline
\end{tabular}
\vspace*{1cm}
\caption{\label{xsec_tablog10beta}Values of the differential cross sections with respect to
  $\beta$.  Other details as in the caption of Table~\ref{xsec_tabQ2}.}
\vspace*{1cm}
\end{center}
\end{table}

\begin{table}[p]
\begin{center}

\begin {tabular}{|r@{,$\;$}l|r@{}l|l r@{}l|p{2.0cm}|p{2.0cm}|r@{}l|}
\hline
\multicolumn{2}{|c|}{$\xpomobs$ bin} & \multicolumn{2}{|c|}{$\xpomobs \, d\sigma/d\xpomobs$}   & \multicolumn{3}{|c|}{$\delta_{{\rm stat}}$} & $\delta_{{\rm syst}}$ & $\delta_{{\rm ES}}$ & \multicolumn{2}{|c|}{$\Delta_{{\rm DIFFR}}$}\\
\multicolumn{2}{|c|}{($\times\,10^{-2}$)} & \multicolumn{2}{|c|}{(pb)}  & \multicolumn{3}{|c|}{(pb)} & (pb) & (pb) & \multicolumn{2}{|c|}{(pb)}\\
\hline  \hline
0.25 & 0.50 & 24.&3 & $\pm$& 1.&8 & $\;^{+0.8}_{-1.5}$ &  $\;^{+1.0}_{-1.1}$ & -0.&5\\
\hline
0.50 & 0.79 & 93$\;$& & $\pm$& 5$\;$& & $\;^{+1}_{-1}$ &  $\;^{+4}_{-5}$ & 0$\;$&\\
\hline
0.79 & 1.26 & 195$\;$& & $\pm$& 7$\;$& & $\;^{+3}_{-9}$ &  $\;^{+9}_{-10}$ & 2$\;$&\\
\hline
1.26 & 1.99 & 306$\;$& & $\pm$& 10$\;$& & $\;^{+10}_{-25}$ &  $\;^{+17}_{-17}$ & 1$\;$&\\
\hline
1.99 & 3.00 & 409$\;$& & $\pm$& 13$\;$& & $\;^{+33}_{-33}$ &  $\;^{+35}_{-30}$ & 26$\;$&\\

\hline
\hline
\end{tabular}
\vspace*{1cm}
\caption{\label{xsec_tablog10xpom}Values of the differential cross sections with respect to
  $\xpomobs$.  Other details as in the caption of Table~\ref{xsec_tabQ2}.}
\vspace*{1cm}

\begin {tabular}{|r@{,$\;$}l|r@{.}l|l r@{.}l|p{2.0cm}|p{2.0cm}|r@{.}l|}
\hline
\multicolumn{2}{|c|}{$\etjj$ bin} & \multicolumn{2}{|c|}{$d\sigma/d\etjj$} & \multicolumn{3}{|c|}{$\delta_{{\rm stat}}$} & $\delta_{{\rm syst}}$ & $\delta_{{\rm ES}}$ &  \multicolumn{2}{|c|}{$\Delta_{{\rm DIFFR}}$} \\
\multicolumn{2}{|c|}{$(\gev)$} & \multicolumn{2}{|c|}{(pb/\gev)}  & \multicolumn{3}{|c|}{(pb/\gev)} & (pb/\gev) & (pb/\gev) &  \multicolumn{2}{|c|}{(pb/\gev)}\\
\hline  \hline
4&5.5 & 51&7 & $\pm$& 1&4 & $\;^{+3.3}_{-3.7}$ &  $\;^{+2.6}_{-2.9}$ & 2&9\\
\hline
5.5 &7.5 & 39&8 & $\pm$& 1&1 & $\;^{+2.6}_{-2.8}$ &  $\;^{+2.2}_{-2.0}$ & 1&8\\
\hline
7.5 & 9.5 & 9&7 & $\pm$& 0&3 & $\;^{+0.7}_{-0.9}$ &  $\;^{+0.8}_{-0.9}$ & 0&2\\
\hline
9.5 & 11.5 & 2&3 & $\pm$& 0&1 & $\;^{+0.1}_{-0.1}$ &  $\;^{+0.3}_{-0.2}$ & 0&1\\
\hline
11.5 & 13.5 & 0&65 & $\pm$& 0&06 & $\;^{+0.03}_{-0.01}$ &  $\;^{+0.08}_{-0.11}$ & 0&03\\
\hline
13.5 & 16 & 0&11 & $\pm$& 0&02 & $\;^{+0.02}_{-0.02}$ &  $\;^{+0.01}_{-0.03}$ & 0&00\\
\hline
\hline
\end{tabular}
\vspace*{1cm}
\caption{\label{xsec_tabetjj}Values of the differential cross sections with respect to
  \etjj.  Other details as in the caption of Table~\ref{xsec_tabQ2}.}
\vspace*{1cm}

\end{center}
\end{table}

\begin{table}[p]
\begin{center}

\begin {tabular}{|r@{,$\;$}l|r@{.}l|l r@{.}l|p{2.0cm}|p{2.0cm}|r@{.}l|}
\hline
\multicolumn{2}{|c|}{$\etajj$ bin} & \multicolumn{2}{|c|}{$d\sigma/d\etajj$}  &  \multicolumn{3}{|c|}{$\delta_{{\rm stat}}$} & $\delta_{{\rm syst}}$ & $\delta_{{\rm ES}}$ &  \multicolumn{2}{|c|}{$\Delta_{{\rm DIFFR}}$}\\
\multicolumn{2}{|c|}{$\;$} & \multicolumn{2}{|c|}{(pb)}  &  \multicolumn{3}{|c|}{(pb)} & (pb) & (pb) &  \multicolumn{2}{|c|}{(pb)}\\
\hline  \hline
-3.5 &-3.0 & 56&6 & $\pm$& 1&9 & $\;^{+2.5}_{-3.8}$ &  $\;^{+7.8}_{-7.6}$ & 1&6\\
\hline
-3.0 &-2.5 & 98&8 & $\pm$& 2&9 & $\;^{+3.6}_{-6.2}$ &  $\;^{+7.2}_{-7.1}$ & 1&8\\
\hline
-2.5 &-2.0 & 89&6 & $\pm$& 2&6 & $\;^{+5.7}_{-6.0}$ &  $\;^{+5.1}_{-4.9}$ & 4&8\\
\hline
-2.0 &-1.5 & 66&1 & $\pm$& 2&1 & $\;^{+4.1}_{-4.2}$ &  $\;^{+3.7}_{-4.1}$ & 3&4\\
\hline
-1.5 &-1.0 & 35&2 & $\pm$& 1&3 & $\;^{+3.3}_{-2.6}$ &  $\;^{+2.7}_{-2.0}$ & 3&0\\
\hline
-1.0 &-0.5 & 13&2 & $\pm$& 0&7 & $\;^{+1.4}_{-1.3}$ &  $\;^{+1.3}_{-1.3}$ & 1&1\\
\hline
-0.5 &$\;$0.0 & 2&1 & $\pm$& 0&2 & $\;^{+0.4}_{-0.5}$ &  $\;^{+0.4}_{-0.3}$ & -0&2\\
\hline
\hline
\end{tabular}
\vspace*{1cm}
\caption{\label{xsec_tabetajj}Values of the differential cross sections with respect to
  \etajj. Other details as in the caption of Table~\ref{xsec_tabQ2}.}
\vspace*{1cm}

\begin {tabular}{|r@{,$\;$}l|r@{.}l|l r@{.}l|p{2.0cm}|p{2.0cm}|r@{.}l|}
\hline
\multicolumn{2}{|c|}{$\zpomobs$} & \multicolumn{2}{|c|}{$d\sigma/d\zpomobs$} &\multicolumn{3}{|c|}{$\delta_{{\rm stat}}$} & $\delta_{{\rm syst}}$ & $\delta_{{\rm ES}}$ & \multicolumn{2}{|c|}{$\Delta_{{\rm DIFFR}}$}\\
\multicolumn{2}{|c|}{$\;$} & \multicolumn{2}{|c|}{(pb)} &\multicolumn{3}{|c|}{(pb)} & (pb) & (pb) & \multicolumn{2}{|c|}{(pb)}\\
\hline  \hline
0 & 0.125 & 24&5 & $\pm$&2&0 & $\;^{+0.9}_{-2.3}$ &  $\;^{+3.5}_{-2.0}$ & -1&9\\
\hline
0.125 & 0.25 & 134&6 & $\pm$&5&4 & $\;^{+8.8}_{-13.8}$ &  $\;^{+12.3}_{-10.8}$ & 0&7\\
\hline
0.25 & 0.375 & 155&1 & $\pm$&5&7 & $\;^{+9.6}_{-12.5}$ &  $\;^{+10.1}_{-9.8}$ & 5&2\\
\hline
0.375 & 0.5 & 133&7 & $\pm$&5&1 & $\;^{+8.3}_{-10.2}$ &  $\;^{+6.1}_{-8.1}$ & 5&7\\
\hline
0.5 & 0.625 & 100&6 & $\pm$&4&2 & $\;^{+5.8}_{-7.8}$ &  $\;^{+5.2}_{-5.4}$ & 2&4\\
\hline
0.625 & 0.75 & 80&4 & $\pm$&3&6 & $\;^{+1.3}_{-2.8}$ &  $\;^{+3.8}_{-3.8}$ & 0&5\\
\hline
0.75 & 0.875 & 55&5 & $\pm$&2&8 & $\;^{+1.7}_{-3.1}$ &  $\;^{+2.8}_{-2.8}$ & -1&5\\
\hline
0.875 & 1 & 31&5 & $\pm$&2&1 & $\;^{+3.1}_{-4.0}$ &  $\;^{+2.2}_{-1.5}$ & -1&3\\
\hline
\hline
\end{tabular}
\vspace*{1cm}
\caption{\label{xsec_tabzpom}Values of the differential cross sections with respect to
  $\zpomobs$. Other details as in the caption of Table~\ref{xsec_tabQ2}.}
\vspace*{1cm}
\end{center}
\end{table}

\begin{table}[p]
\begin{center}
\begin {tabular}{|r@{,$\;$}l|r@{.}l|l r@{.}l|p{2.0cm}|p{2.0cm}|r@{.}l|}
\hline
\multicolumn{2}{|c|}{$\xgammaobs$ bin} & \multicolumn{2}{|c|}{$d\sigma/d\xgammaobs$}  & \multicolumn{3}{|c|}{$\delta_{{\rm stat}}$} & $\delta_{{\rm syst}}$ & $\delta_{{\rm ES}}$ &  \multicolumn{2}{|c|}{$\Delta_{{\rm DIFFR}}$}\\
\multicolumn{2}{|c|}{$\;$} & \multicolumn{2}{|c|}{(pb)}  & \multicolumn{3}{|c|}{(pb)} & (pb) & (pb) &  \multicolumn{2}{|c|}{(pb)}\\
\hline  \hline
0 & 0.25 & 5&3 & $\pm$&0&6 & $\;^{+0.8}_{-0.5}$ &  $\;^{+0.6}_{-0.2}$ & 0&7\\
\hline
0.25 & 0.5 & 25&0 & $\pm$&1&3 & $\;^{+2.8}_{-1.5}$ &  $\;^{+1.5}_{-2.2}$ & 2&6\\
\hline
0.5 & 0.75 & 87&4 & $\pm$&3&1 & $\;^{+4.5}_{-5.7}$ &  $\;^{+7.8}_{-8.2}$ & 3&5\\
\hline
0.75 & 1 & 240&5 & $\pm$&6&7 & $\;^{+11.0}_{-17.2}$ &  $\;^{+12.4}_{-11.7}$ & 5&9\\
\hline
\hline
\end{tabular}
\vspace*{1cm}
\caption{\label{xsec_tabxgamma}Values of the differential cross sections with respect to
  \xgammaobs.  Other details as in the caption of Table~\ref{xsec_tabQ2}.}
\vspace*{1cm}
\end{center}
\end{table}

%
%
%

\begin{table}[p]
\begin{center}
\begin {tabular}{|r@{,$\;$}l|r@{.}l|l r@{.}l|p{2.0cm}|p{2.0cm}|r@{.}l|}
\hline
\multicolumn{2}{|c|}{$\zpomobs$ bin}& \multicolumn{2}{|c|}{$d\sigma/d\zpomobs dE^{*}_{T,j1}$} & \multicolumn{3}{|c|}{$\delta_{{\rm stat}}$} & $\delta_{{\rm syst}}$ & $\delta_{{\rm ES}}$ &  \multicolumn{2}{|c|}{$\Delta_{{\rm DIFFR}}$}\\
\multicolumn{2}{|c|}{$\;$}& \multicolumn{2}{|c|}{(pb/\gev)}  & \multicolumn{3}{|c|}{(pb/\gev)} & (pb/\gev) & (pb/\gev) & \multicolumn{2}{|c|}{(pb/\gev)}\\
\hline  \hline
\multicolumn{11}{|c|}{$5.0< E^{*}_{T,j1}<6.5 \gev (<E^{*}_{T,j1}>=5.7 \gev)$}\\

\hline
0 & 0.25 & 31&9 & $\pm$&1&5 & $\;^{+2.1}_{-4.2}$ &  $\;^{+1.9}_{-1.4}$ & -1&2\\
\hline
0.25 & 0.375 & 53&1 & $\pm$&2&6 & $\;^{+4.2}_{-4.6}$ &  $\;^{+2.3}_{-2.3}$ & 3&0\\
\hline
0.375 & 0.5 & 46&7 & $\pm$&2&4 & $\;^{+3.0}_{-3.0}$ &  $\;^{+1.9}_{-2.4}$ & 2&5\\
\hline
0.5 & 0.625 & 35&3 & $\pm$&2&1 & $\;^{+1.1}_{-1.7}$ &  $\;^{+1.6}_{-1.8}$ & 1&0\\
\hline
0.625 & 0.75 & 29&3 & $\pm$&1&9 & $\;^{+0.3}_{-1.9}$ &  $\;^{+1.3}_{-1.2}$ & -0&8\\
\hline
0.75 & 0.875 & 18&4 & $\pm$&1&4 & $\;^{+1.2}_{-1.9}$ &  $\;^{+0.8}_{-0.8}$ & -1&0\\
\hline
0.875 & 1 & 11&4 & $\pm$&1&2 & $\;^{+0.3}_{-1.0}$ &  $\;^{+0.5}_{-0.5}$ & -0&5\\
\hline




\multicolumn{11}{|c|}{$6.5< E^{*}_{T,j1}<8.0 \gev (<E^{*}_{T,j1}>=7.2 \gev)$}\\
\hline
0 & 0.25 & 13&2 & $\pm$&0&5 & $\;^{+1.4}_{-1.5}$ &  $\;^{+0.9}_{-0.6}$ & 0&17\\
\hline
0 & 0.25 & 13&2 & $\pm$&0&8 & $\;^{+1.4}_{-1.5}$ &  $\;^{+0.9}_{-0.6}$ & 0&2\\
\hline
0.25 & 0.375 & 25&9 & $\pm$&1&5 & $\;^{+1.2}_{-2.2}$ &  $\;^{+1.9}_{-2.0}$ & -0&9\\
\hline
0.375 & 0.5 & 21&9 & $\pm$&1&3 & $\;^{+1.9}_{-1.8}$ &  $\;^{+1.6}_{-0.9}$ & 1&9\\
\hline
0.5 & 0.625 & 18&3 & $\pm$&1&2 & $\;^{+0.8}_{-1.0}$ &  $\;^{+1.2}_{-1.0}$ & 0&7\\
\hline
0.625 & 0.75 & 14&8 & $\pm$&1&1 & $\;^{+0.8}_{-0.9}$ &  $\;^{+0.6}_{-0.6}$ & 0&2\\
\hline
0.75 & 0.875 & 12&4 & $\pm$&1&0 & $\;^{+0.8}_{-0.9}$ &  $\;^{+0.6}_{-0.8}$ & 0&0\\
\hline
0.875 & 1 & 5&6 & $\pm$&0&7 & $\;^{+0.2}_{-0.2}$ &  $\;^{+0.5}_{-0.2}$ & -0&1\\
\hline
\multicolumn{11}{|c|}{$8.0< E^{*}_{T,j1}<16.0 \gev (<E^{*}_{T,j1}>=9.7 \gev)$}\\
\hline
0.25 & 0.375 & 2&4 & $\pm$&0&2 & $\;^{+0.4}_{-0.3}$ &  $\;^{+0.2}_{-0.2}$ & 0&2\\
\hline
0.375 & 0.5 & 2&4 & $\pm$&0&2 & $\;^{+0.2}_{-0.2}$ &  $\;^{+0.2}_{-0.3}$ & -0&1\\
\hline
0.5 & 0.625 & 1&9 & $\pm$&0&1 & $\;^{+0.1}_{-0.2}$ &  $\;^{+0.2}_{-0.2}$ & 0&0\\
\hline
0.625 & 0.75 & 1&7 & $\pm$&0&1 & $\;^{+0.1}_{-0.0}$ &  $\;^{+0.2}_{-0.2}$ & 0&1\\
\hline
0.75 & 0.875 & 1&4 & $\pm$&0&1 & $\;^{+0.0}_{-0.1}$ &  $\;^{+0.1}_{-0.1}$ & 0&0\\
\hline
0.875 & 1 & 0&80 & $\pm$&0&09 & $\;^{+0.01}_{-0.10}$ &  $\;^{+0.10}_{-0.08}$ & -0&05\\
\hline
\hline
\end{tabular}
\vspace*{1cm}
\caption{\label{xsec_tabzpomDD}Values of the double differential cross
  sections with respect to $\zpomobs$ in bins of $E^{*}_{T,{\rm j1}}$. Other details as in the caption of Table~\ref{xsec_tabQ2}. }
\end{center}
\end{table}

\begin{table}[p]
\begin{center}
\end{center}
\end{table}

\begin{table}[p]
\begin{center}
\begin {tabular}{|r@{,$\;$}l|r@{.}l|l r@{.}l|p{2.0cm}|p{2.0cm}| r@{.}l|}
\hline
\multicolumn{2}{|c|}{$\zpomobs$ bin} &\multicolumn{2}{|c|}{ $d\sigma/d\zpomobs dQ^{2}$}&\multicolumn{3}{|c|}{ $\delta_{{\rm stat}}$} & $\delta_{{\rm syst}}$ & $\delta_{{\rm ES}}$& \multicolumn{2}{|c|}{$\Delta_{{\rm DIFFR}}$}\\
\multicolumn{2}{|c|}{$\;$} &\multicolumn{2}{|c|}{ $({\rm pb}/\gev^{2})$}&\multicolumn{3}{|c|}{$ ({\rm pb}/\gev^{2})$} &  $({\rm pb}/\gev^{2})$ & $({\rm pb}/\gev^{2})$ &\multicolumn{2}{|c|}{$({\rm pb}/\gev^{2})$}\\
\hline  \hline
\multicolumn{11}{|c|}{$5 < Q^{2} <12 \gev^{2} (<Q^{2}>=8.1 \gev^{2})$}\\
\hline
0 & 0.25 & 5&1 & $\pm$&0&3 & $\;^{+0.4}_{-0.5}$ &  $\;^{+0.4}_{-0.4}$ & 0&0\\
\hline
0.25 & 0.375 & 8&7 & $\pm$&0&4 & $\;^{+0.6}_{-0.7}$ &  $\;^{+0.7}_{-0.7}$ & -0&1\\
\hline
0.375 & 0.5 & 7&2 & $\pm$&0&4 & $\;^{+0.4}_{-0.5}$ &  $\;^{+0.5}_{-0.5}$ & 0&3\\
\hline
0.5 & 0.625 & 5&2 & $\pm$&0&3 & $\;^{+0.1}_{-0.2}$ &  $\;^{+0.4}_{-0.4}$ & 0&1\\
\hline
0.625 & 0.75 & 4&3 & $\pm$&0&3 & $\;^{+0.1}_{-0.2}$ &  $\;^{+0.3}_{-0.2}$ & 0&1\\
\hline
0.75 & 0.875 & 2&9 & $\pm$&0&2 & $\;^{+0.1}_{-0.1}$ &  $\;^{+0.2}_{-0.2}$ & -0&1\\
\hline
0.875 & 1 & 1&5 & $\pm$&0&2 & $\;^{+0.1}_{-0.2}$ &  $\;^{+0.1}_{-0.1}$ & -0&1\\
\hline




\multicolumn{11}{|c|}{$12 < Q^{2} <25 \gev^{2} (<Q^{2}>=17.2 \gev^{2})$}\\
\hline
0 & 0.25 & 1&43 & $\pm$&0&09 & $\;^{+0.14}_{-0.16}$ &  $\;^{+0.09}_{-0.07}$ & 0&08\\
\hline
0.25 & 0.375 & 3&0 & $\pm$&0&2 & $\;^{+0.4}_{-0.2}$ &  $\;^{+0.1}_{-0.2}$ & 0&4\\
\hline
0.375 & 0.5 & 2&3 & $\pm$&0&1 & $\;^{+0.1}_{-0.2}$ &  $\;^{+0.1}_{-0.1}$ & 0&1\\
\hline
0.5 & 0.625 & 2&0 & $\pm$&0&1 & $\;^{+0.1}_{-0.2}$ &  $\;^{+0.1}_{-0.1}$ & 0&0\\
\hline
0.625 & 0.75 & 1&6 & $\pm$&0&1 & $\;^{+0.1}_{-0.1}$ &  $\;^{+0.1}_{-0.1}$ & 0&0\\
\hline
0.75 & 0.875 & 1&2 & $\pm$&0&1 & $\;^{+0.0}_{-0.1}$ &  $\;^{+0.1}_{-0.1}$ & -0&1\\
\hline
0.875 & 1 & 0&61 & $\pm$&0&07 & $\;^{+0.01}_{-0.03}$ &  $\;^{+0.04}_{-0.03}$ & -0&01\\
\hline
\end{tabular}
\vspace*{1cm}
\caption{\label{xsec_tabzpomDD1}Values of the double differential cross
  sections with respect to $\zpomobs$ in bins of $E^{*}_{T,{\rm j1}}$. Other details as in the caption of Table~\ref{xsec_tabQ2}. }
\end{center}
\end{table}

\begin{table}[p]
\begin{center}
\begin {tabular}{|r@{,$\;$}l|r@{.}l|l r@{.}l|p{2.0cm}|p{2.0cm}|r@{.}l|}
\hline
\multicolumn{2}{|c|}{$\zpomobs$ bin}& \multicolumn{2}{|c|}{$d\sigma/d\zpomobs dQ^{2}$} &\multicolumn{3}{|c|}{ $\delta_{{\rm stat}}$} & $\delta_{{\rm syst}}$& $\delta_{{\rm ES}}$ &  \multicolumn{2}{|c|}{$\Delta_{{\rm DIFFR}}$}\\
\multicolumn{2}{|c|}{$\;$}& \multicolumn{2}{|c|}{$({\rm pb}/\gev^{2})$ } &\multicolumn{3}{|c|}{$({\rm pb}/\gev^{2})$} & $({\rm pb}/\gev^{2})$ & $({\rm pb}/\gev^{2})$ &  \multicolumn{2}{|c|}{$({\rm pb}/\gev^{2})$}\\
\hline  \hline
\multicolumn{11}{|c|}{$25 < Q^{2} <50 \gev^{2} (<Q^{2}>=35.2 \gev^{2})$}\\
\hline
0 & 0.25 & 0&51 & $\pm$&0&04 & $\;^{+0.03}_{-0.08}$ &  $\;^{+0.03}_{-0.02}$ & -0&06\\
\hline
0.25 & 0.375 & 1&03 & $\pm$&0&07 & $\;^{+0.07}_{-0.13}$ &  $\;^{+0.06}_{-0.07}$ & -0&03\\
\hline
0.375 & 0.5 & 1&00 & $\pm$&0&07 & $\;^{+0.06}_{-0.06}$ &  $\;^{+0.04}_{-0.06}$ & 0&05\\
\hline
0.5 & 0.625 & 0&77 & $\pm$&0&06 & $\;^{+0.06}_{-0.02}$ &  $\;^{+0.04}_{-0.04}$ & 0&06\\
\hline
0.625 & 0.75 & 0&60 & $\pm$&0&05 & $\;^{+0.03}_{-0.04}$ &  $\;^{+0.04}_{-0.03}$ & -0&01\\
\hline
0.75 & 0.875 & 0&44 & $\pm$&0&04 & $\;^{+0.01}_{-0.02}$ &  $\;^{+0.03}_{-0.04}$ & 0&00\\
\hline
0.875 & 1 & 0&24 & $\pm$&0&03 & $\;^{+0.01}_{-0.01}$ &  $\;^{+0.02}_{-0.01}$ & 0&00\\
\hline
\multicolumn{11}{|c|}{$50 < Q^{2} <100 \gev^{2} (<Q^{2}>=69.5 \gev^{2})$}\\
\hline
0 & 0.25 & 0&10 & $\pm$&0&01 & $\;^{+0.00}_{-0.01}$ &  $\;^{+0.01}_{-0.01}$ & 0&00\\
\hline
0.25 & 0.375 & 0&25 & $\pm$&0&02 & $\;^{+0.05}_{-0.01}$ &  $\;^{+0.02}_{-0.01}$ & 0&05\\
\hline
0.375 & 0.5 & 0&28 & $\pm$&0&03 & $\;^{+0.02}_{-0.01}$ &  $\;^{+0.01}_{-0.02}$ & 0&02\\
\hline
0.5 & 0.625 & 0&20 & $\pm$&0&02 & $\;^{+0.02}_{-0.01}$ &  $\;^{+0.01}_{-0.01}$ & 0&02\\
\hline
0.625 & 0.75 & 0&16 & $\pm$&0&02 & $\;^{+0.01}_{-0.00}$ &  $\;^{+0.01}_{-0.01}$ & 0&01\\
\hline
0.75 & 0.875 & 0&13 & $\pm$&0&02 & $\;^{+0.00}_{-0.00}$ &  $\;^{+0.01}_{-0.01}$ & 0&00\\
\hline
0.875 & 1 & 0&11 & $\pm$&0&02 & $\;^{+0.01}_{-0.02}$ &  $\;^{+0.01}_{-0.01}$ & -0&01\\
\hline
\end{tabular}
\vspace*{1cm}
\caption{\label{xsec_tabzpomDD2}Values of the double differential cross
  sections with respect to $\zpomobs$ in bins of $E^{*}_{T,{\rm j1}}$. Other details as in the caption of Table~\ref{xsec_tabQ2}. }
\end{center}
\end{table}



\vspace*{4cm}

%
%

%


%
%
\end{document}